\documentclass[a4paper]{article}
\pdfoutput=1

\usepackage[utf8]{inputenc}
\usepackage[T1]{fontenc}
\usepackage[english]{babel}
\usepackage{fullpage}
\usepackage[pdftex,breaklinks=true,bookmarks=true,pdfborder={0 0 0},colorlinks=false]{hyperref}

\usepackage{listings}
\lstdefinelanguage{sl}{alsolanguage=C,morekeywords={sl_def,sl_enddef,sl_index,sl_parm,sl_getp,sl_setp,__asm__,sl_glparm,sl__static,sl_shparm,__volatile__,__typeof__,sl_create,sl_createsync,sl_glarg,sl_sharg,sl_geta,sl_sync,sl__exclusive,sl__static,sl_detach,sl_decl,sl_typedef_fptr,sl_glfparm,sl_glfarg,sl_shfparm,sl_shfarg,inline,sl_seta}}

\lstdefinelanguage{hs}{morekeywords={where,do,import,hiding}}

\usepackage{cleveref}
\usepackage{graphicx}
\usepackage[caption=false]{subfig}

\newcommand{\ie}{i.e.~}
\newcommand{\eg}{e.g.~}

\usepackage{enumitem}
\setlist{nolistsep}
\usepackage{array}

\begin{document}

\author{Raphael ‘kena’ Poss}
\title{Machines are benchmarked by code, \\ not algorithms}

\maketitle

\begin{abstract}
  This article highlights how small modifications to either the source
  code of a benchmark program or the compilation options may impact
  its behavior on a specific machine. It argues that for evaluating
  machines, benchmark providers and users be careful to ensure
  reproducibility of results based on the machine code actually
  running on the hardware and not just source code. The article uses
  color to grayscale conversion of digital images as a running
  example.
\end{abstract}

\setcounter{tocdepth}{2}
\tableofcontents

\clearpage
\section{Introduction}

In computer science, hardware providers use benchmarks to demonstrate
their technology. A benchmark suite should be composed so that each of the
\emph{benchmarks} exercises features of the technology in a way that
is representative of a larger class of programs.  Two desirable
properties oppose each other when designing a suite: it should be as
\emph{concise} as possible to minimize the time and effort necessary
to run; and as \emph{complete} as possible so as to evaluate the most
features.

Conciseness is easy to measure, even without understanding the
suite. In contrast, measuring completeness requires
understanding the technology and how the suite exercises its
features. To facilitate this, practitioners usually
\emph{characterize} individual benchmarks using general traits, and
describe separately how groups of features in the technology are
collectively exercised by each trait. For example, a
benchmark can be said to be ``memory intensive'', with the general
understanding that memory intensive programs will exercise the memory
subsystem of the computer more than the processor or its I/O
facilities.

Meanwhile, another reality of computer science is automatic program
transformation. Benchmarks are rarely expressed by their authors in
the same language used by the machine to run them. Compilers, and
particularly optimization algorithms in compilers, can radically
change the structure of a program, depending on configuration or from
one version of a compiler to another.
This reality entails the main argument of this article: \emph{machine
features are ultimately exercised by the machine code of benchmark
programs, not their source code.}

The following corollaries should serve as words of caution for
newcomers to the ``art'' of designing and using benchmarks:

\begin{enumerate}
\item a single benchmark program can test different machine features
  depending on which compiler and optimization options are being used;
\item small code changes to a benchmark that keep the overall function
  of an algorithm identical
  can  change the set of machine features the program is
  exercising, depending on compiler and optimizations;
\item it is not sufficient to describe the algorithm or the source
  code of a benchmark to characterize which features of the technology
  it exercises.
\end{enumerate}

What to do about this? A gentle reminder seems in order:
\begin{enumerate}
\item do not claim directly which machine features a program exercises
  unless you know the machine code being run;
\item do not use different compilers or optimization settings than
  those recommended by benchmark providers unless you understand how
  your choice will impact the machine behavior;
\item do not change a benchmark source code until you understand how
  changes impact the machine behavior after compilation;
\item if the program code implicity depends on third-party components
  at run-time (\eg operating system services), do not use different
  components than those recommended by the benchmark providers unless
  you understand how your choice will impact the machine behavior;
\item when presenting results, ensure \emph{both} full
  \emph{reproducibility} of the results and \emph{access} to the
  program sources, compiler sources, operating software and platform specification so that the
  audience can cross-check the results.
\end{enumerate}

Reproducibility, eg. by making the machine code available, is a
general requirement of the scientific method; access to sources is
specific to computer science and made necessary by the sheer
complexity of duplicating platforms and reverse-engineering machine
code to understand what it does.

The rest of this article illustrates this point throughout a running
example from digital imaging.

\section{Example}\label{sec:ex}

A common operation in photography is the conversion of colored
pictures into ``black\&white'' or shade of grays.
With digital images, a discretization of the image into pixels is
assumed. An image is then characterized by a discrete \emph{domain} of
pixel coordinates, and the \emph{value} at each position in the
domain.
The
``color to grayscale conversion'' can be generally described for
arbitrary domains using a pixel-to-pixel numerical conversion, for example:
$$ rgbtogray_\mathrm{pixel}(r, g, b) = 0.3 \times r + 0.59 \times g + 0.11
\times b$$

This specific formula accounts for the greater sensitivity of the
human eye for green and blue relative to red.  Given this pixel
conversion function, the conversion of an entire image can then be
defined as the application of the function to all pixels in the
domain. This general definition is already suitable as an
\emph{algorithm} in a declarative functional language.  However, this
specification alone is not sufficient to define a benchmark because it
does not define any \emph{work}: a dataset to be operated on, with a
specific workload in terms of image size, value types, data structure
for pixels, etc.

\subsection{Work definitions}

In general, a \emph{work definition} includes at least:
\begin{itemize}
\item the \emph{program implementation}, defining both the computation
  algorithm(s) and the plumbing code that ``connects'' the
  algorithms to the environment;
\item a description of the \emph{program interface} of the benchmark
  which indicates to the user how to run the program, what input it
  accepts and what output it produces; and
\item a set of reference \emph{workloads} that the benchmark should be
  run with in order to produce comparable results.
\end{itemize}.

An example work definition can be phrased as follows.
\begin{itemize}
\item implementation: the computation
algorithm and I/O code are given in Haskell in \cref{lst:r2g}.
The concrete data structures and the strategy to perform the
computation are provided by the Repa and DevIL libraries;
\item interface: the input and output file names are given as
first and second positional arguments on the command line.  Although
this specific implementation happens to accept any input image format
accepted by the DevIL library, the benchmark interface only mandates
binary PPM input and binary PGM output;
\item workloads: the benchmark should be run on input images of
  the sizes given in \cref{tab:szs}. These sizes have been chosen
  to span a logarithmic range from 10 to 200M pixels, with multiples
  of both binary and decimal powers.
\end{itemize}

\subsection{Example result report}

\begin{figure}
\centering
\includegraphics[width=\textwidth]{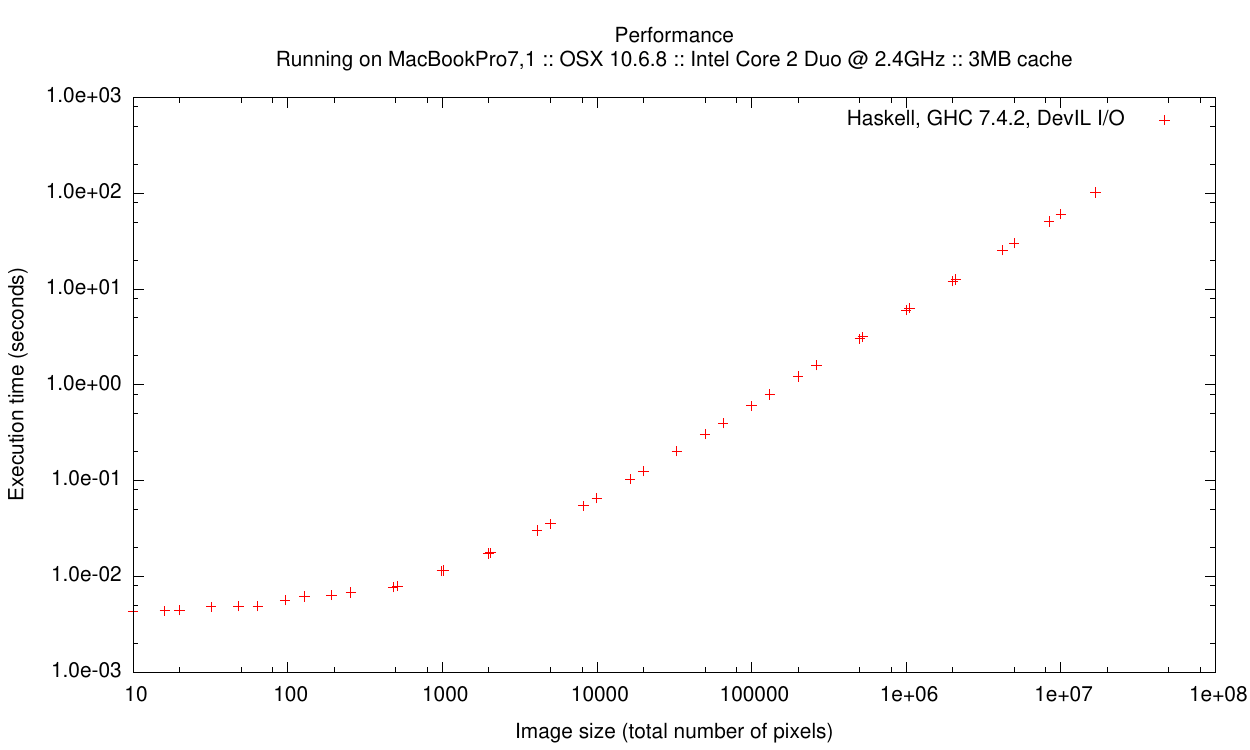}
\caption{Example performance of the program in \cref{lst:r2g} with image sizes from \cref{tab:szs}.}\label{fig:hs1}
\end{figure}

The performance of this benchmark on the author's workstation is
pictured in \cref{fig:hs1}. This graph reports the minimum
execution time from five runs for each of the input sizes in
\cref{tab:szs}. The title of the graph reports the platform where the
program was run.

\begin{figure}
\centering
\includegraphics[width=\textwidth]{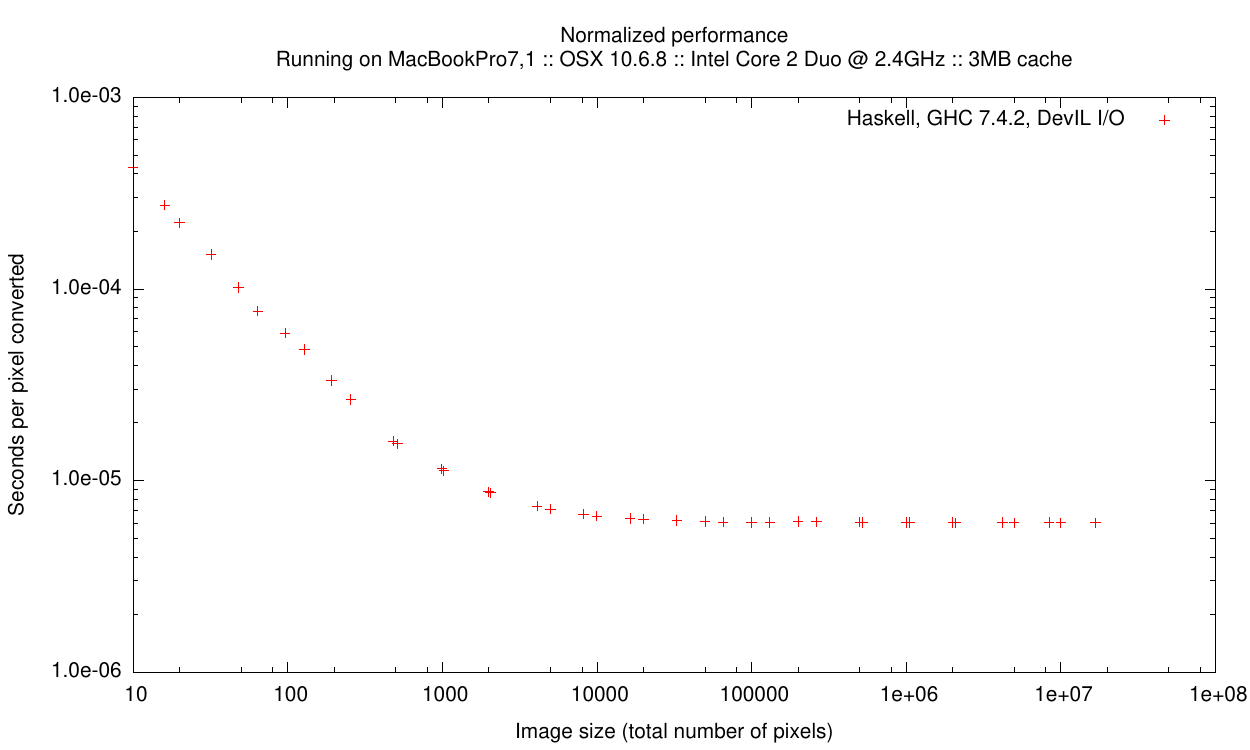}
\caption{Same results as \cref{fig:hs1}, normalized by input size.}\label{fig:hs2}
\end{figure}

Since the execution time is expected to be linear in its
input size, a more direct impression of the performance can be obtained by
normalizing, \ie dividing the observed execution time by the input
size. As the result in \cref{fig:hs2} shows, this program's
performance increases with the input size until approximately size
10000, beyond which the performance stays stable at around 6$\mu$s/pixel.

\subsection{Language and operating software}

Both the programming language, its compiler and the surrounding
operating software impact the performance of a given algorithm.

Consider for example the alternate C implementation of the previous
benchmark, given in \cref{lst:cr2g,lst:cr2g:d2d}. This can be compiled
with either \cref{lst:ppm:il} to use the same DevIL library as
the Haskell code from \cref{lst:r2g} for image I/O, or using
\cref{lst:ppm} to use POSIX system calls for I/O instead.

\begin{figure}
\centering
\includegraphics[width=\textwidth]{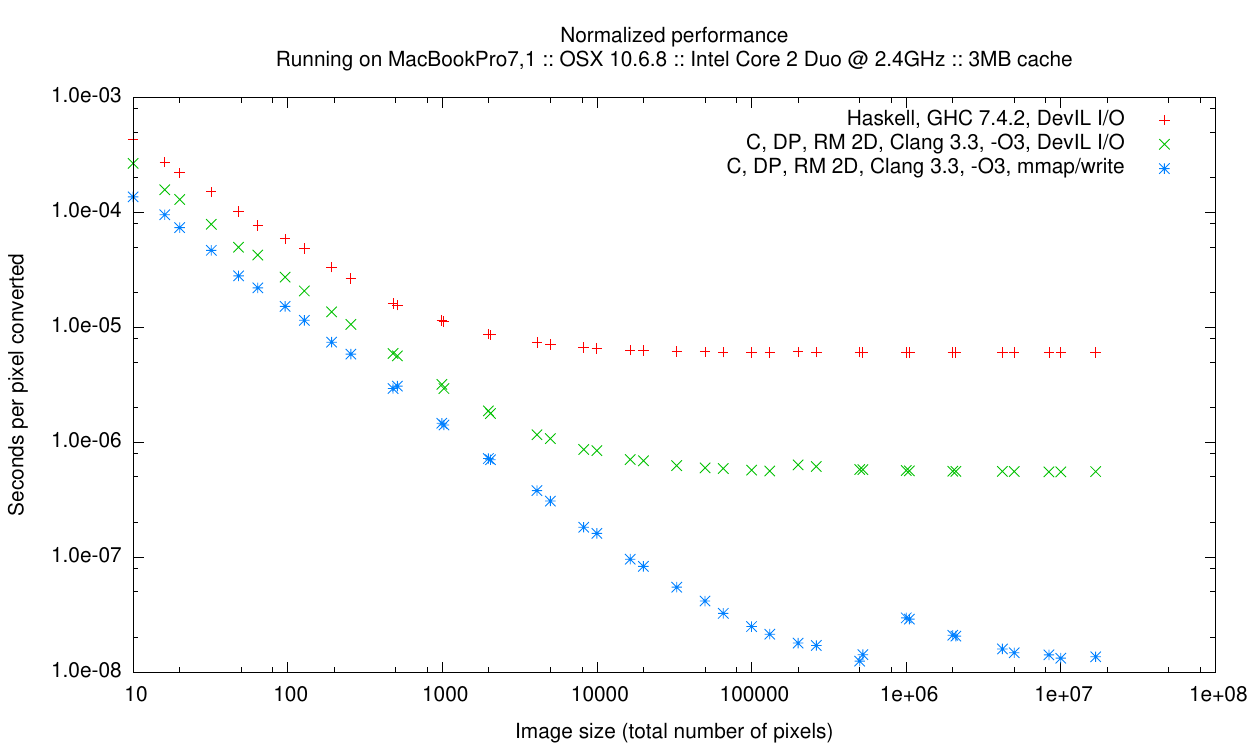}
\caption{Relative performance of different languages and operating software.}\label{fig:lang}
\end{figure}

As the results in \cref{fig:lang} show, there are large performance
differences depending on which software stack is used. In this
example, the C implementation using the same image I/O routines
(DevIL) is about 10$\times$ faster than the Haskell
implementation. When using the same C code but the POSIX I/O routines
instead of DevIL, another gain of 40$\times$ is possible.

These differences are commonly called \emph{abstraction overheads},
and are frequently accepted by programmers as the necessary price to
pay to benefit from the higher-level languages and operating software
services.

\subsection{I/O overheads}

Benchmark can be broadly classified into purely computational benchmarks,
which only measure the performance of a program assuming all the data
manipulated is already present ``within the system'', and system
benchmarks which also include the time to move data in and out of the
system. The program presented above belong to the latter category, because the
measured time to execute includes the time to load and store images
from and to files.

\begin{figure}
\centering
\includegraphics[width=\textwidth]{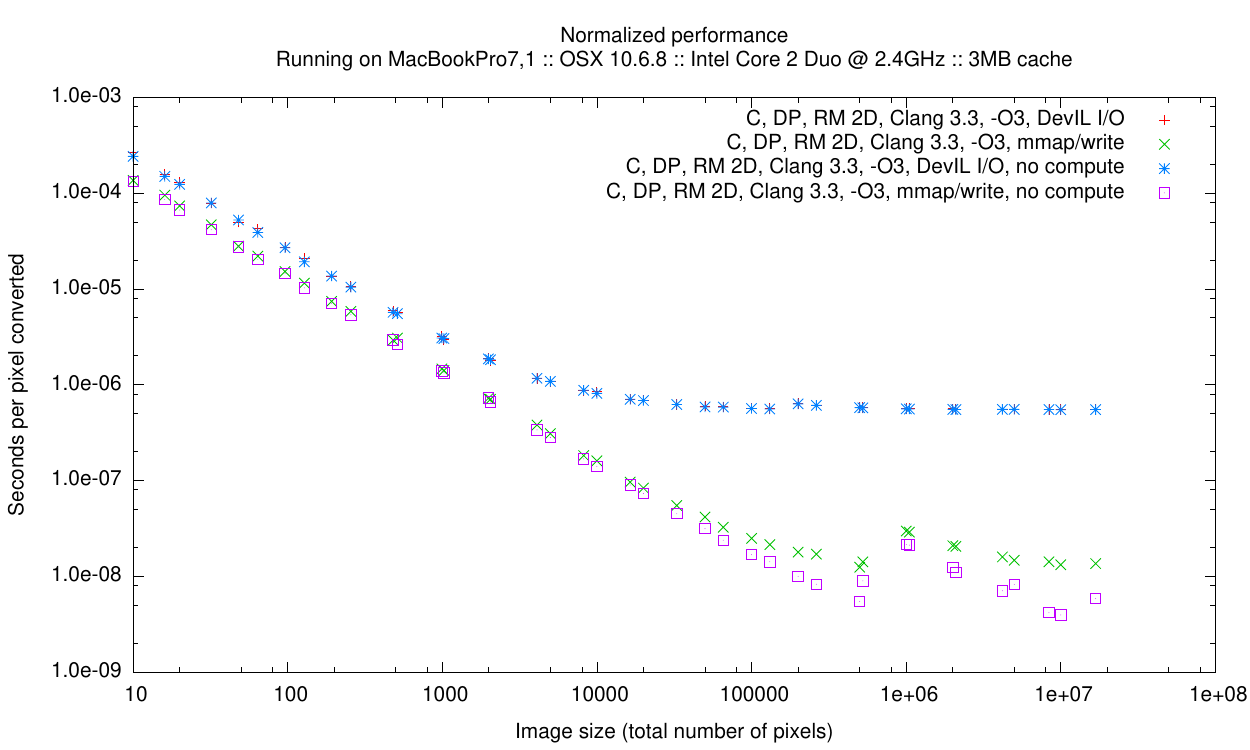}
\caption{Relative performance of different I/O methods without computation.}\label{fig:empty}
\end{figure}

\begin{figure}
\centering
\includegraphics[width=\textwidth]{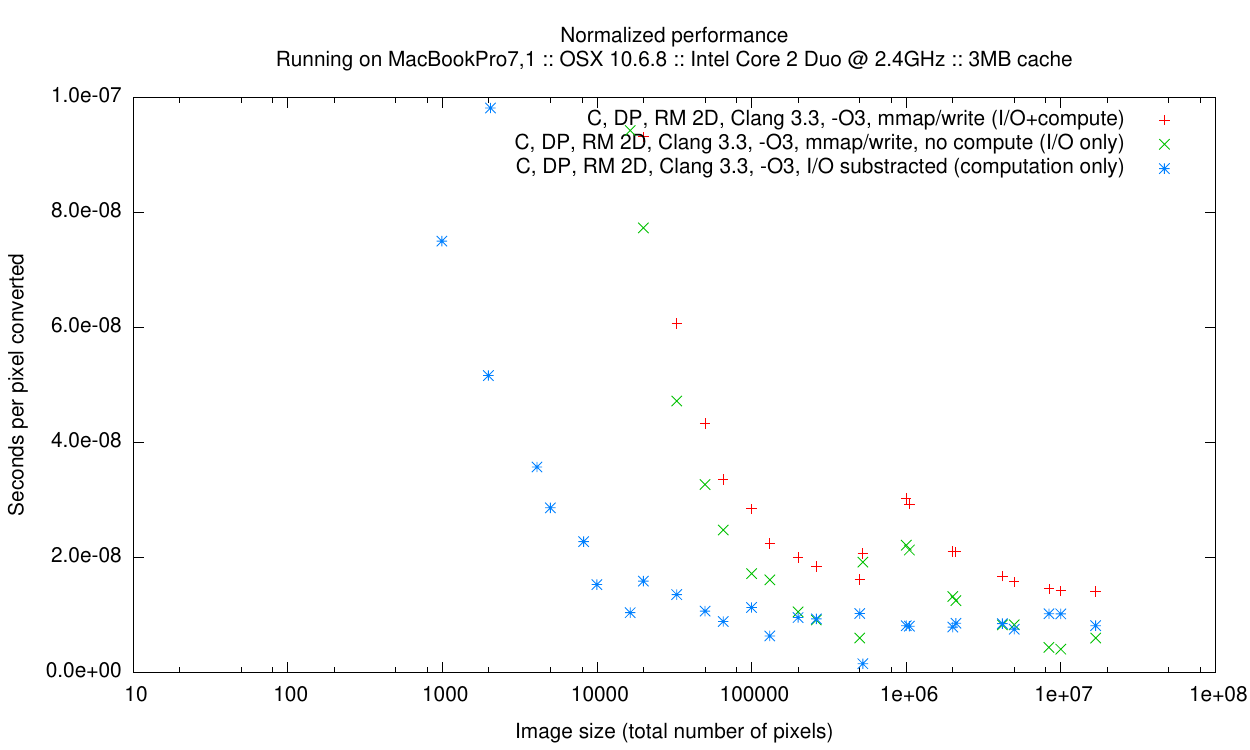}
\caption{Separate I/O and computation performance (POSIX I/O).}\label{fig:sub}
\end{figure}

\begin{figure}
\centering
\includegraphics[width=\textwidth]{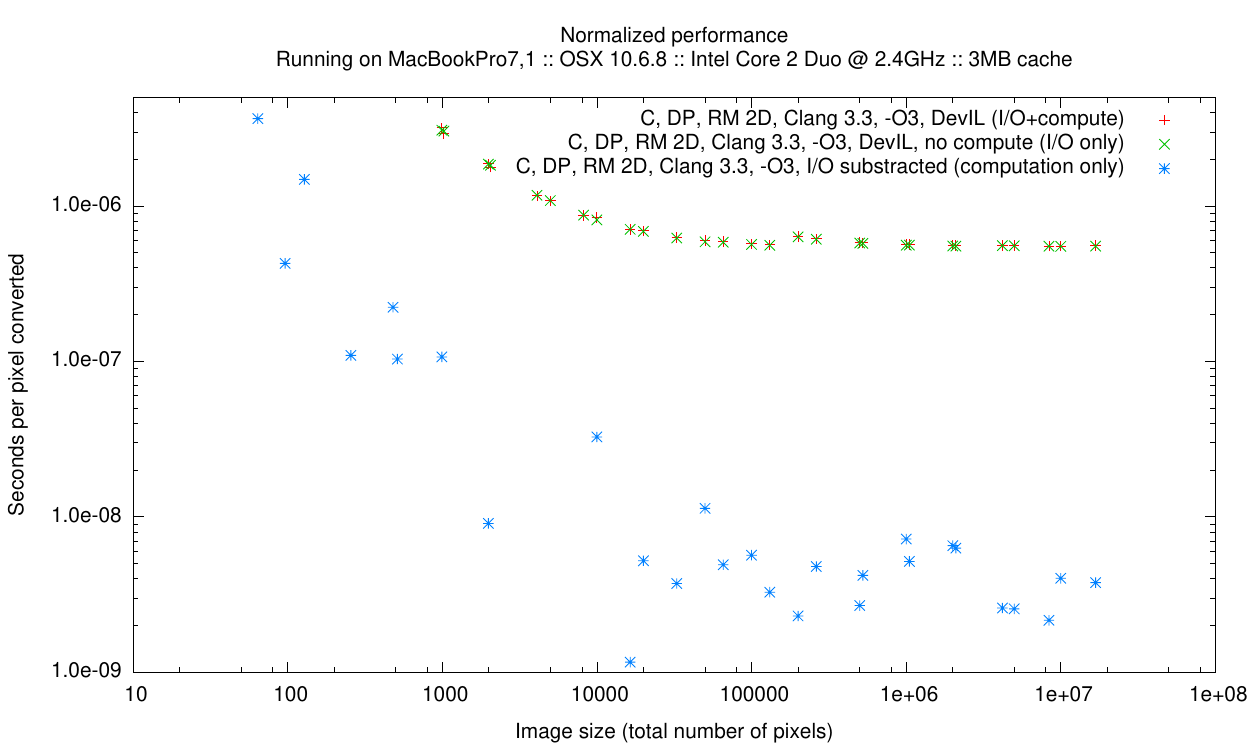}
\caption{Separate I/O and computation performance (DevIL I/O).}\label{fig:sub2}
\end{figure}

To gain more insight into the behavior of a machine, it is usually
desirable, when possible, to separate the I/O from the computation
costs. We can do this with the C implementation from \cref{lst:cr2g}
by removing the body of the computation, leaving only the I/O
code. By doing so, we can measure the I/O time in isolation, as
demonstrated in \cref{fig:empty}. Once the I/O time is known, it is
possible to isolate the computation time by substracting the I/O time
from the total time.

An example of this separation is given in \cref{fig:sub,fig:sub2}. As
these figures illustrate, when the I/O time dominates
(\cref{fig:sub2}) it becomes difficult to extract a meaningful
impression of the computation time in isolation. Meanwhile, as shown
with \cref{fig:sub}, the I/O time dominates as long as the working set
fits in the cache; when the computation becomes memory-intensive the
I/O time becomes comparatively smaller.

\subsection{Traversal order}

\begin{figure}
\centering
\subfloat[Overall performance, I/O
included]{\includegraphics[width=\textwidth]{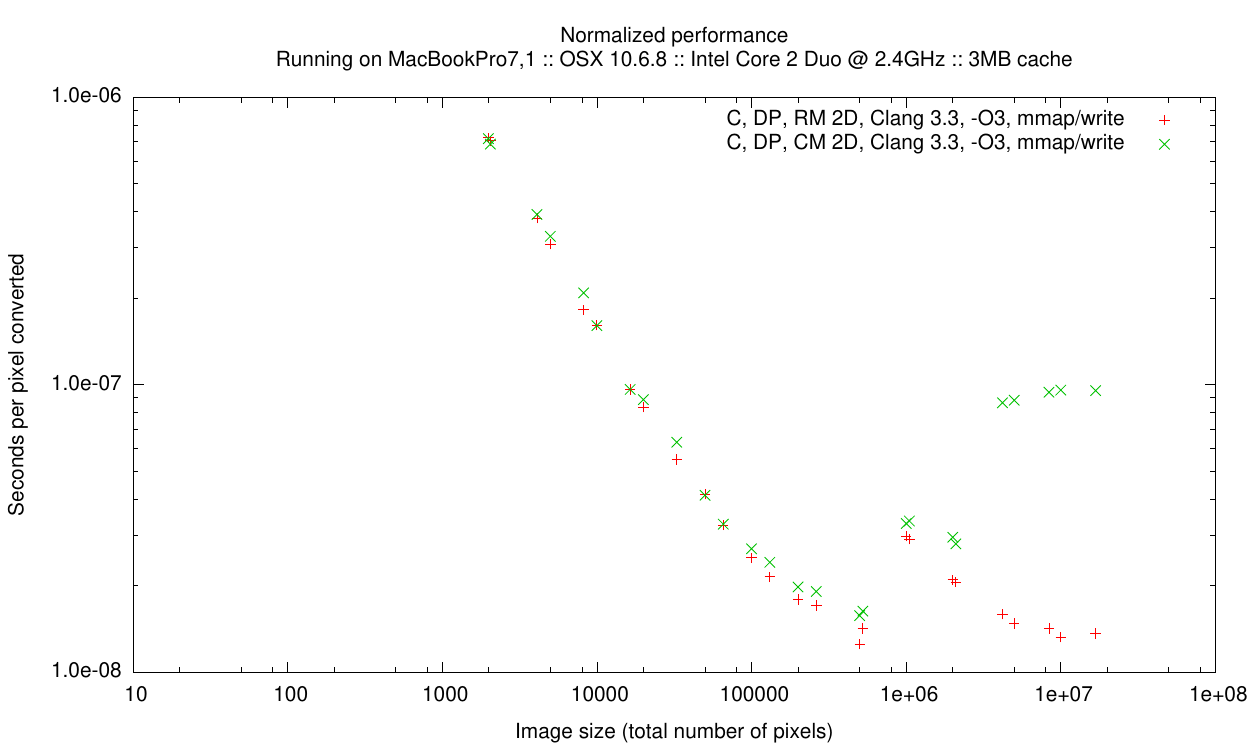}}

\subfloat[Computation performance, I/O excluded]{\includegraphics[width=\textwidth]{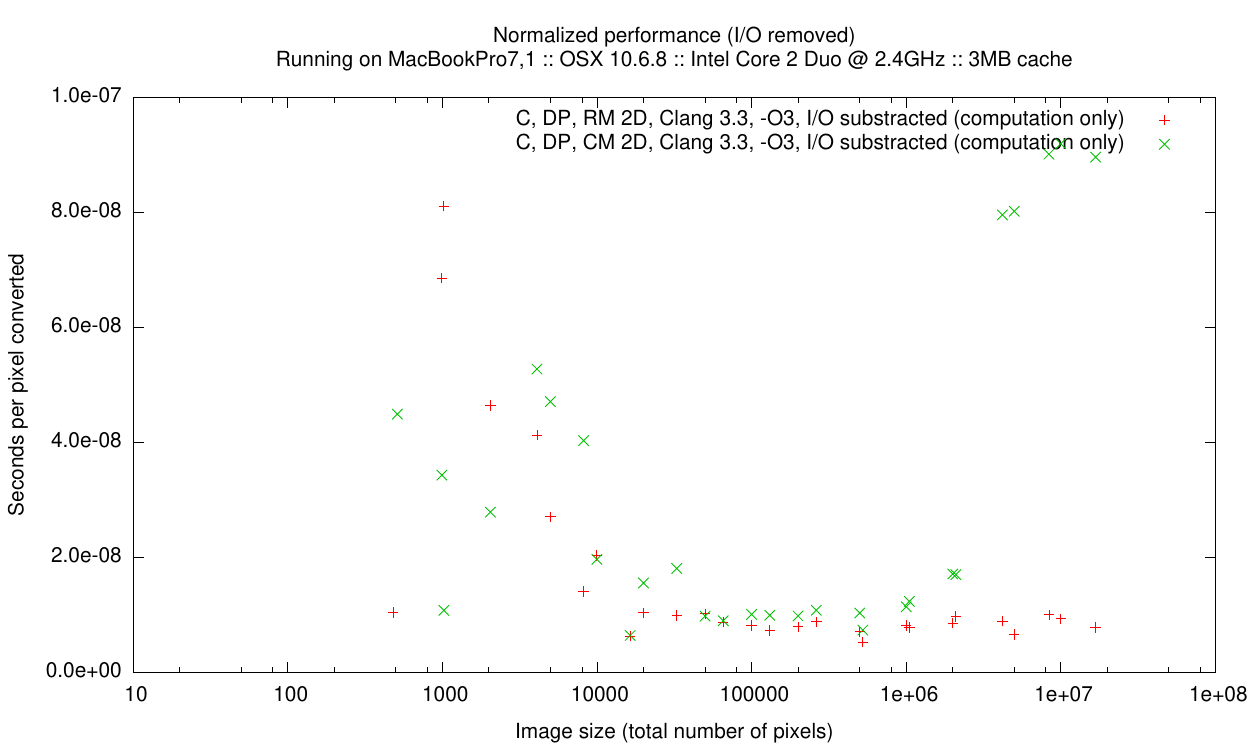}}
\caption{Relative performance of different traversal orders.}\label{fig:trans}
\end{figure}

The order in which an algorithm accesses values
in the machine's memory may impact its performance.
Consider for example the difference between \cref{lst:cr2g:d2d} and
\cref{lst:cr2g:d2dt}. \Cref{lst:cr2g:d2d} accesses the data in
row-major order (RM), \ie successive inner iterations access
contiguous locations in memory. \Cref{lst:cr2g:d2dt} accesses the data
in column-major order (CM), \ie successive inner iterations access
items in memory separated by a full row.
Example results are given in \cref{fig:trans}.

As this figure shows, caches can hide the impact of different
traversal orders as long as the working set fits entirely in
cache. When the working set becomes larger than the cache, the
column-major order causes more cache misses and the performance
degrades. Conversely, it is possible to deduce the overall cache size
of a chip by observing at which working set size the performance
of different traversal order diverges.

In this situation, the column-major traversal can be said to be more
memory-intensive than the row-major traversal, and thus excercises the
cache system more.

\subsection{Loop nesting}

Control flow loops are the means in sequential code to apply an
operation to the elements of an array. If the array is
multi-dimensional, the straightforward encoding uses nested
loops (\cref{lst:cr2g:d2d}). However, if the array data is contiguous
in memory, it is usually possible to flatten nested loops to
a single level, as in \cref{lst:cr2g:d1d}.

\begin{figure}
\centering
\subfloat[Overall performance, I/O included.]{\includegraphics[width=\textwidth]{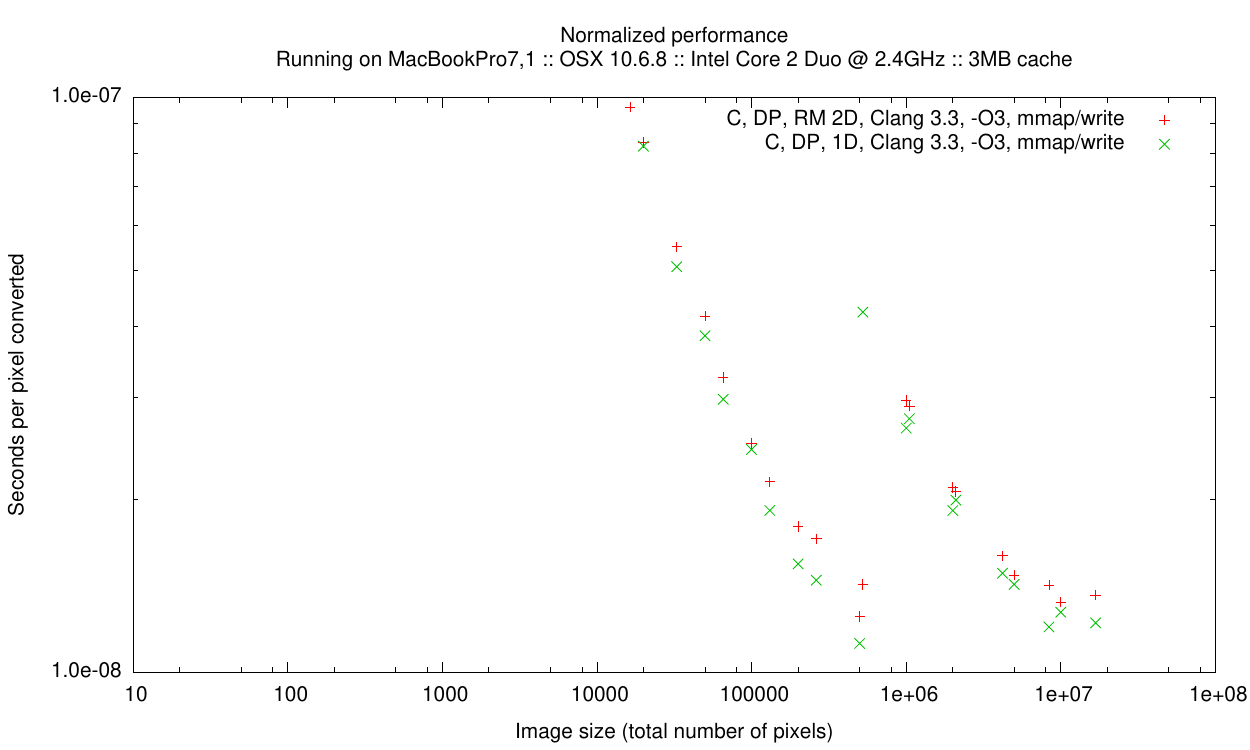}}

\subfloat[Computation performance, I/O excluded.]{\includegraphics[width=\textwidth]{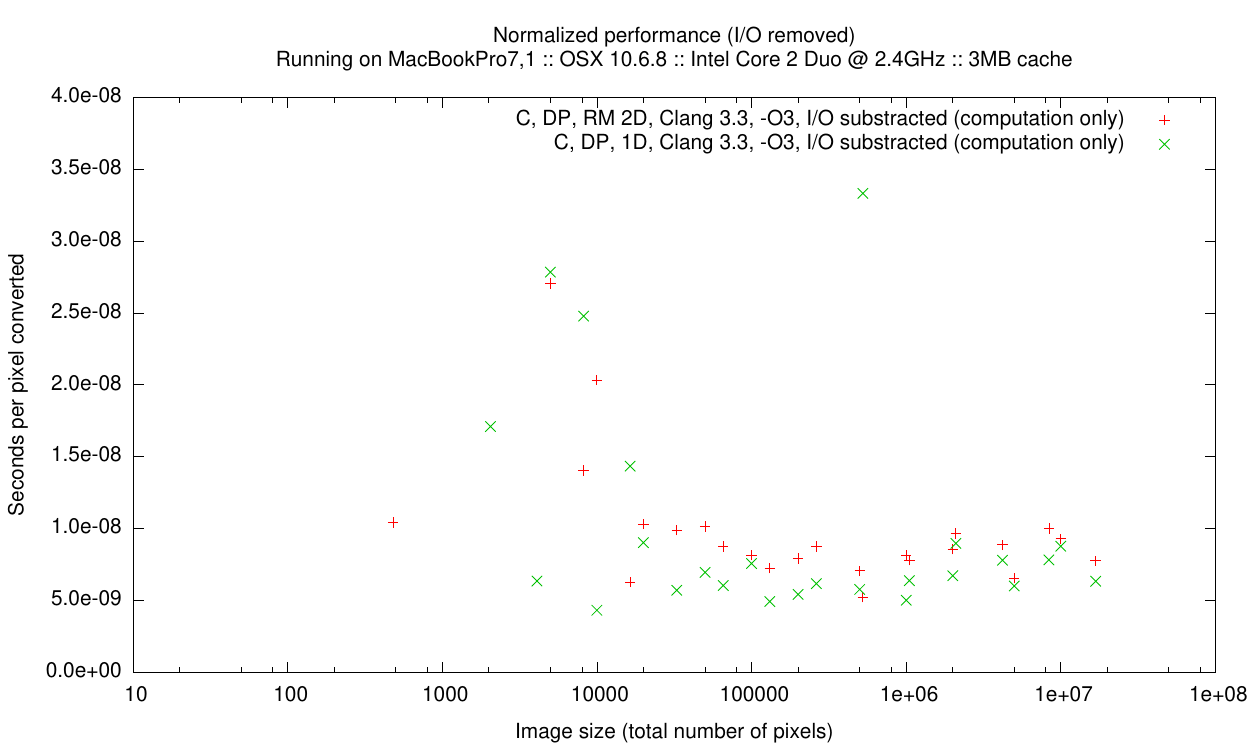}}
\caption{Relative performance of nested and flattened loops.}\label{fig:dim}
\end{figure}

Although the resulting machine code performs the same arithmetic
operations, and the total number of inner iterations stays the same,
flattening can impact performance as shown in \cref{fig:dim}. In this
example, the flattened loop (1D) performs up to 13\% faster than the
nested loop (2D).

This difference can be explained by either a simpler
index computation or a lower number of branch prediction misses. In
other words, the nested loop implementation exercises more the integer
arithmetic unit and the branch prediction / tolerance facilities of
the processor than the flattened loop implementation.

\subsection{Data types and operation widths}

Any numerical computation is translated to concrete machine code that
operates on specific data types. A type determines both the
representation of values as arrangements of bits, and the concrete
implementation of the operations on these values.
For a given computational problem, there may be multiple
possible choices of types that would yield acceptable numerical
results. Each of these possible types, in turn, translates
to different machine code and thus different machine behavior.

\begin{figure}
\centering
\subfloat[Overall performance, I/O included.]{\includegraphics[width=\textwidth]{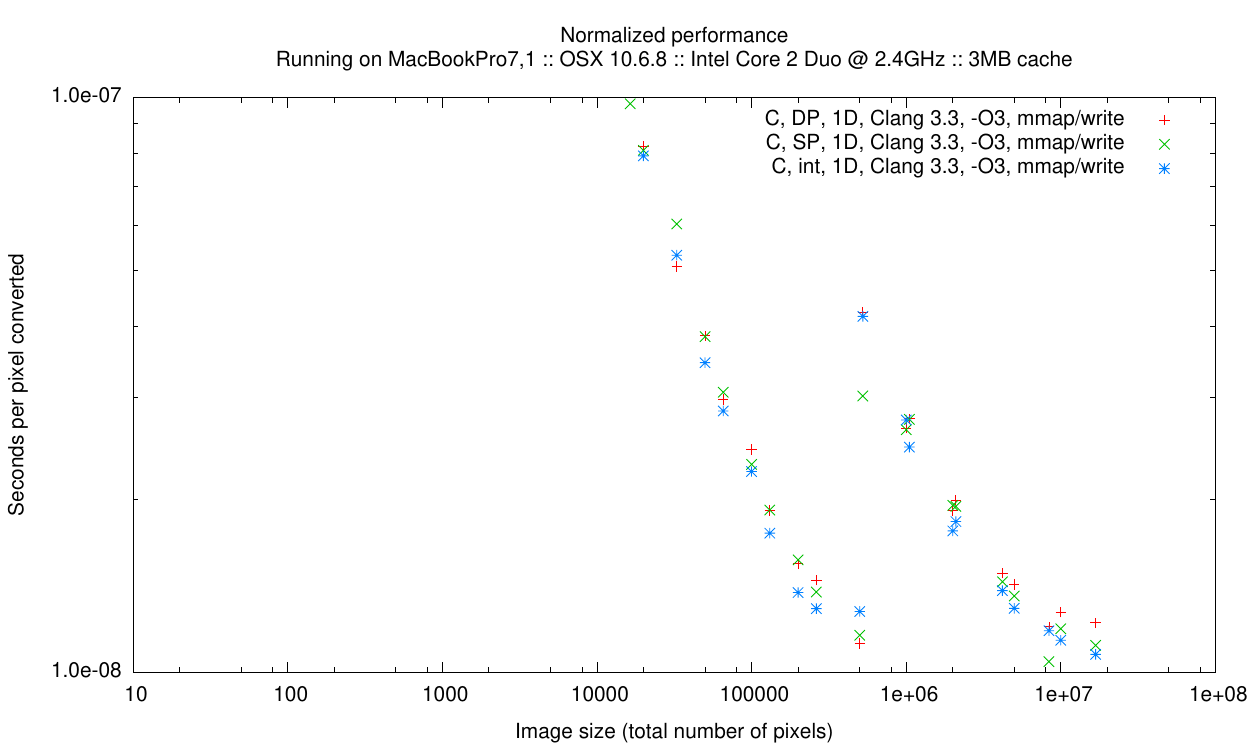}}

\subfloat[Computation performance, I/O excluded.]{\includegraphics[width=\textwidth]{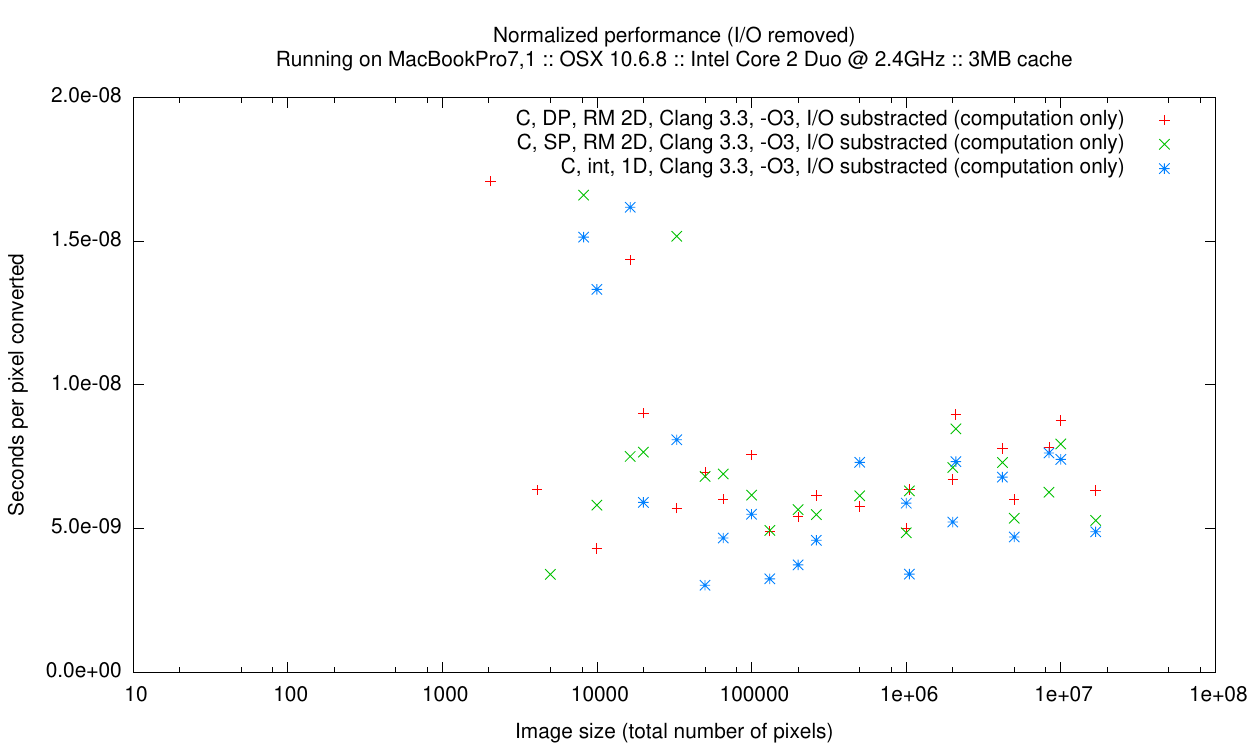}}
\caption{Relative performance of double precision, single precision
  and integer arithmetic.}\label{fig:type}
\end{figure}

For example, the general problem
described at the start of \cref{sec:ex} can be computed using single
precision floating point arithmetic (\cref{lst:cr2g:s1d}) or even
integer arithmetic (\cref{lst:cr2g:i1d}) with no
result differences visible by the human eye. As the results from
\cref{fig:type} show, on this particular machine and for this
particular computational problem, the single precision
performance can be up to 8\%  higher than double precision, and the integer
performance up to 13\% higher.

From the perspective of characterization, the floating-point
implementations exercises more the support for floating point in the
architecture, whereas the integer implementation exercises the integer
unit. With a machine without a hardware FPU, the performance
difference should be expected to be even more dramatic.

\subsection{Compiler}

Even when using the same source code and work definition, the choice
of compiler to transform the code may have an impact on machine
behavior.

\begin{figure}
\centering
\subfloat[Overall performance, I/O
included.]{\includegraphics[width=\textwidth]{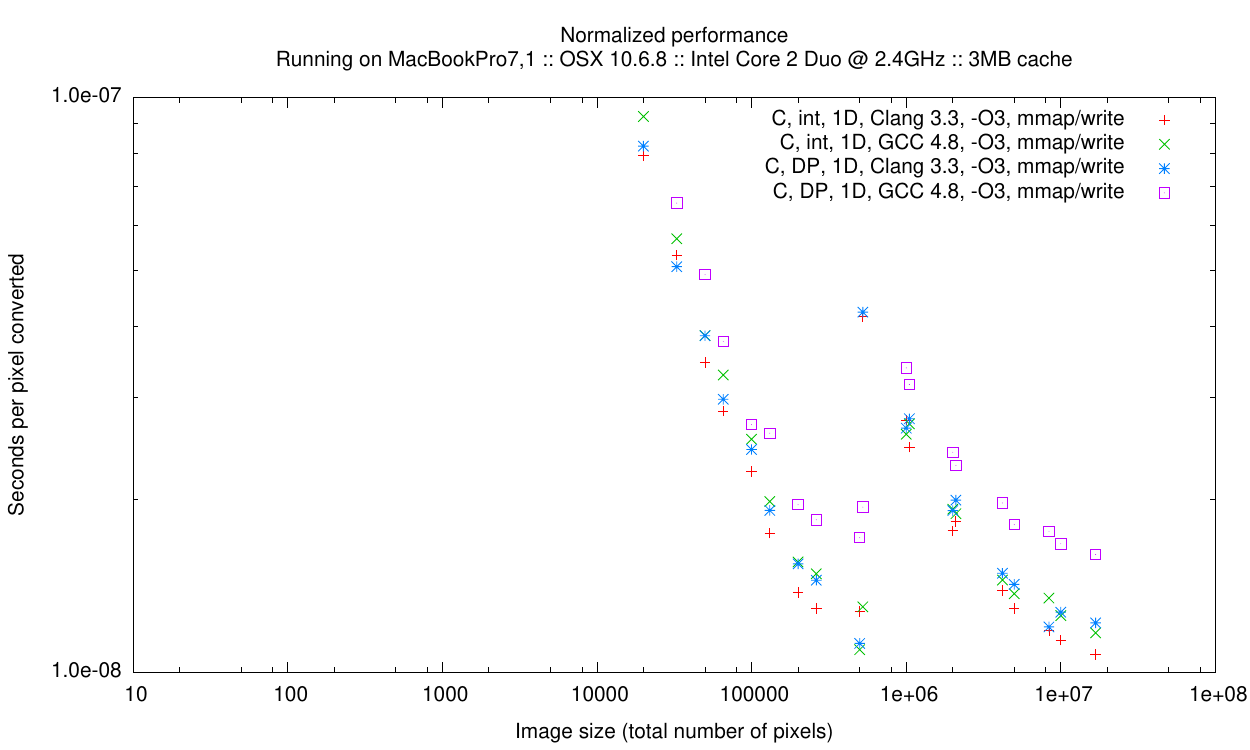}}

\subfloat[Computation performance, I/O excluded.]{\includegraphics[width=\textwidth]{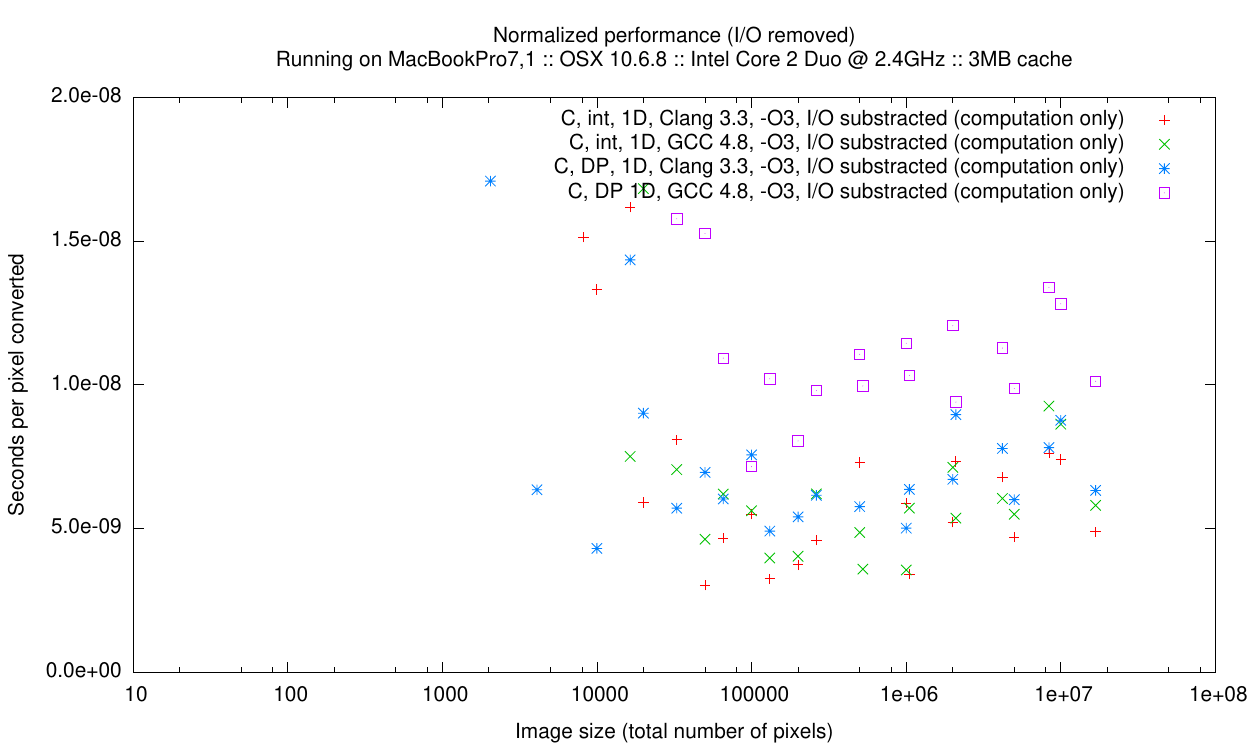}}
\caption{Relative performance of code generated by Clang/LLVM and GNU C.}\label{fig:comp}
\end{figure}

As shown by \cref{fig:comp}, in this particular environment the
machine code generated by Clang/LLVM 3.3 peforms reliably faster then
the code generated by GNU C 4.8 when using double precision floating point
arithmetic; whereas for integer arithmetic there is no clear
performance ordering between the two.

\subsection{Compiler optimizations}

Each compiler can transform the source code into machine code in
different ways depending on which optimizations are enabled, and their parameters.

\begin{figure}
\centering
\subfloat[Overall performance, I/O
included.]{\includegraphics[width=\textwidth]{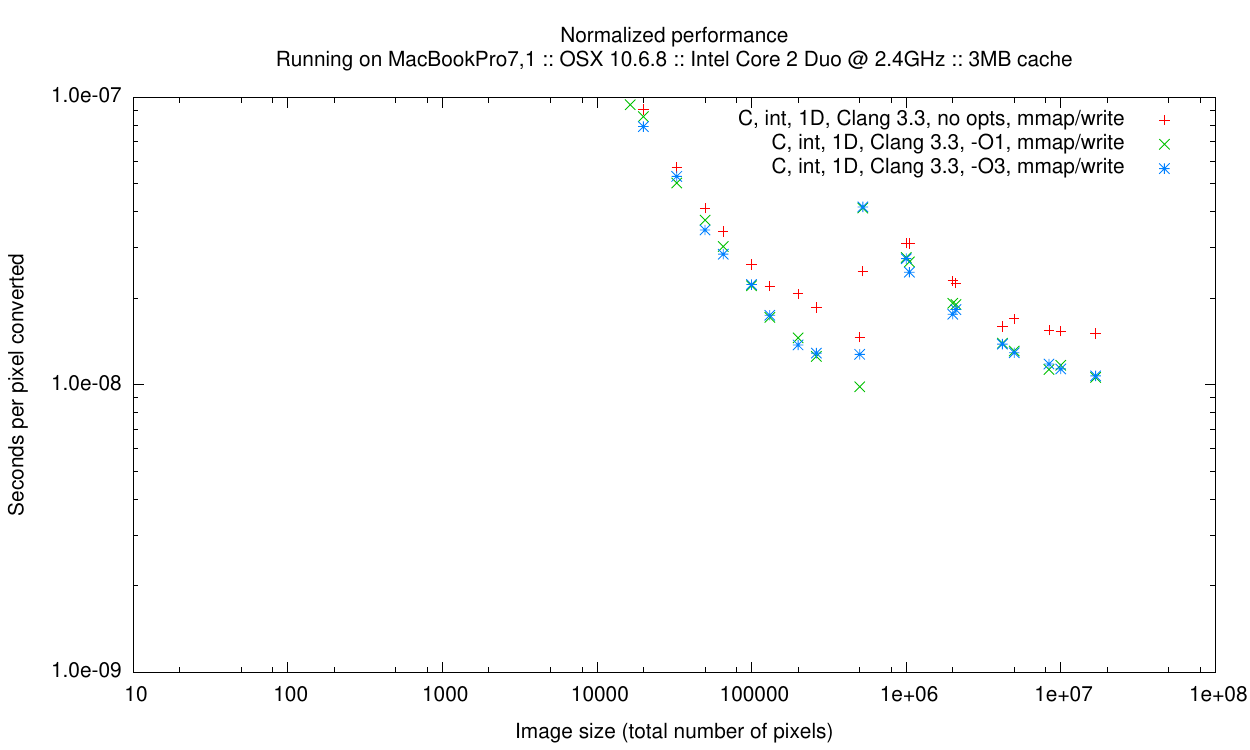}}

\subfloat[Computation performance, I/O excluded.]{\includegraphics[width=\textwidth]{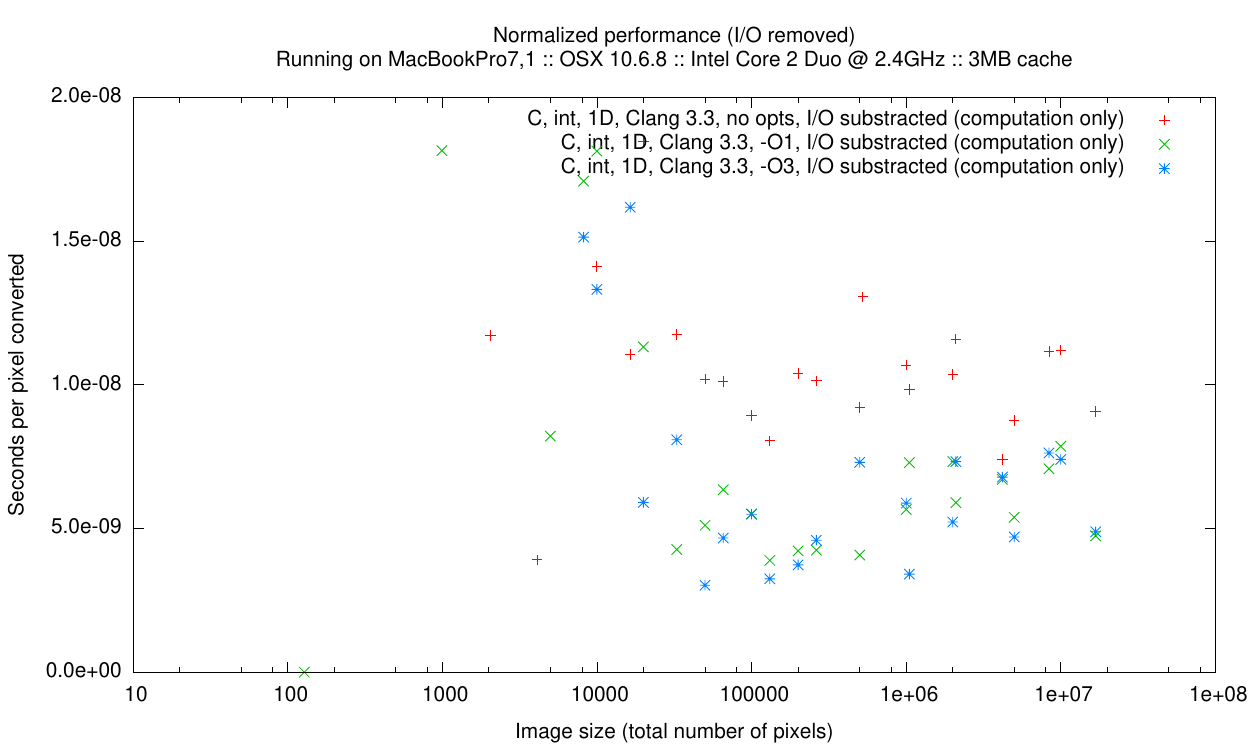}}
\caption{Relative performance of code generated by Clang/LLVM with different
  optimization levels.}\label{fig:flags}
\end{figure}

An example is given in \cref{fig:flags}, where the Clang/LLVM compiler is
configured
with different optimization levels. In this environment, the
optimization level ``\texttt{-O3}'' performs up to 40\% better than
when no optimization level is specified. However, as illustrated in
the graph, for some combinations of input sizes (\eg 1024x488, at
499712 pixels) the optimization level
``\texttt{-O1}'' may actually yield better performance than
``\texttt{-O3}'' (up to 20\% better).

\section{Summary and conclusion}

This article has argued that machine features are ultimately exercised
by the machine code of benchmark programs, not their source
code. Through a running example from digital imaging, we have seen how
the choice of programming language, software tool chain, data layout
and traversal order, data types and operations, compiler and compiler
optimizations can impact the machine behavior with little or no
changes to the benchmark source code.

This presentation is intended as a reminder to practitioners who use
benchmarks to evaluate platforms and machines, that they should carefully report
all the parameters necessary for a peer to reproduce their results.
This report should include details about the tool chain and usage parameters, and not only
the source code of the benchmark programs.

\clearpage

\begin{lstlisting}[float,label=lst:r2g,caption={Example Haskell program that
performs a color to greyscale conversion.},language=hs]
import Data.Array.Repa (computeS, traverse, extent, size)
import Data.Array.Repa.IO.DevIL (readImage, writeImage, runIL, Image(..))
import Data.Array.Repa.Index (Z(..), (:.)(..))
import System.Environment (getArgs)

rgb2gray (r,g,b) = 0.3*r + 0.59*g + 0.11*b

main = do
  [fin, fout] <- getArgs
  (RGB x) <- runIL $ readImage fin
  let y = computeS $ traverse x to2d (cvt rgb2gray)
  runIL $ writeImage fout (Grey y)

to2d (Z:.i:.j:._) = Z:.i:.j
cvt cf f ix = floor $ cf $ rgb f ix
  where
    rgb f ix = (get $ ix:.0, get $ ix:.1, get $ ix:.2)
         where get = fromIntegral . f
\end{lstlisting}

\begin{table}
\centering
\begin{tabular}{ll|ll|ll}
Size & Pixels & Size & Pixels & Size & Pixels \\
\hline
5x2 & 10 &         64x31 & 1984 &     512x512 & 262144 \\
4x4 & 16 &         64x32 & 2048 &     1024x488 & 499712 \\
5x4 & 20 &         64x64 & 4096 &     1024x512 & 524288 \\
8x4 & 32 &         128x39 & 4992 &    1024x976 & 999424 \\
8x6 & 48 &         128x64 & 8192 &    1024x1024 & 1048576  \\
8x8 & 64 &         128x78 & 9984 &    2048x976 & 1998848   \\
16x6 & 96 &        128x128 & 16384 &  2048x1024 & 2097152  \\
16x8 & 128 &       256x78 & 19968 &   2048x2048 & 4194304  \\
16x12 & 192 &      256x128 & 32768 &  2048x2441 & 4999168  \\
16x16 & 256 &      256x195 & 49920 &  4096x2048 & 8388608  \\
32x15 & 480 &      256x256 & 65536 &  4096x2441 & 9998336  \\
32x16 & 512 &      512x195 & 99840 &  4096x4096 & 16777216 \\
32x31 & 992 &      512x256 & 131072 & & \\
32x32 & 1024 &     512x390 & 199680 & &  \\
\end{tabular}
\caption{Example reference sizes for a benchmark input.}\label{tab:szs}
\end{table}

\begin{lstlisting}[float,label=lst:cr2g,caption={Example C program
  that performs a color to greyscale conversion.},escapechar=?]
#include <stdlib.h> // malloc

typedef unsigned char (*rgbimage)[3];
typedef unsigned char *greyimage;

#include "ppm.c" // read_ppm, write_pgm

int main(int argc, char *argv[]) {
    unsigned w, h;  rgbimage x; greyimage y;

    x = read_ppm(argv[1], &w, &h);
    y = malloc(w*h);

#include "compute.c"

    write_pgm(argv[2], y, w, h);
    return 0;
}
\end{lstlisting}

\begin{lstlisting}[float,label=lst:cr2g:d2d,caption={double2d.c, DP, RM 2D: body of the
  computation in \cref{lst:cr2g}.}]
#define RGB2GRAY(R, G, B)   .3 * R + .59 * G + .11 * B

unsigned i, j;
for (j = 0; j < h; ++j)
    for (i = 0; i < w; ++i)
        y[j*w+i] = RGB2GRAY(x[j*w+i][0], x[j*w+i][1], x[j*w+i][2]);
\end{lstlisting}

\begin{lstlisting}[float,label=lst:cr2g:d2dt,caption={double2dt.c, DP, CM 2D: body of the
  computation in \cref{lst:cr2g}.}]
#define RGB2GRAY(R, G, B)   .3 * R + .59 * G + .11 * B;

unsigned i, j;
for (i = 0; i < w; ++i)
    for (j = 0; j < h; ++j)
        y[j*w+i] = RGB2GRAY(x[j*w+i][0], x[j*w+i][1], x[j*w+i][2]);
\end{lstlisting}

\begin{lstlisting}[float,label=lst:cr2g:d1d,caption={double1d.c, DP, 1D: body of the
  computation in \cref{lst:cr2g}.}]
#define RGB2GRAY(R, G, B)   .3 * R + .59 * G + .11 * B

unsigned i;
for (i = 0; i < w*h; ++i)
    y[i] = RGB2GRAY(x[i][0], x[i][1], x[i][2]);
\end{lstlisting}

\begin{lstlisting}[float,label=lst:cr2g:s1d,caption={float1d.c, SP, 1D: body of the
  computation in \cref{lst:cr2g}.}]
#define RGB2GRAY(R, G, B)   .3f * R + .59f * G + .11f * B

unsigned i;
for (i = 0; i < w*h; ++i)
     y[i] = RGB2GRAY(x[i][0], x[i][1], x[i][2]);
\end{lstlisting}

\begin{lstlisting}[float,label=lst:cr2g:i1d,caption={int1d.c, integer, 1D: body of the
  computation in \cref{lst:cr2g}.}]
#define RGB2GRAY(R, G, B)   (30 * R + 59 * G + 11 * B) / 100

unsigned i;
for (i = 0; i < w*h; ++i)
    y[i] = RGB2GRAY(x[i][0], x[i][1], x[i][2]);
\end{lstlisting}

\begin{lstlisting}[float,label=lst:ppm:il,caption={ppm\_il.c: interface to image I/O using the DevIL
  library.}]
#include <IL/il.h> // IL functions

void write_pgm(char *filename, greyimage img, unsigned w, unsigned h) {
    ilTexImage(w, h, 1, 1, IL_LUMINANCE, IL_UNSIGNED_BYTE, img);
    ilSaveImage(filename);
}

rgbimage read_ppm(char *filename, unsigned *w, unsigned *h) {
    ILuint handle;
    ilInit();
    ilGenImages(1, &handle);
    ilBindImage(handle);
    ilLoadImage(filename);
    *w = ilGetInteger(IL_IMAGE_WIDTH);
    *h = ilGetInteger(IL_IMAGE_HEIGHT);
    return (rgbimage)ilGetData();
}
\end{lstlisting}
\begin{lstlisting}[float,label=lst:ppm,caption={ppm\_unix.c: interface to image I/O using POSIX
  functions.},escapechar=?]
#include <stdio.h> // perror,snprintf
#include <unistd.h> // read,write,close
#include <fcntl.h> // open
#include <sys/mman.h> // mmap
#include <stdlib.h> // strtoul,exit

#define SYS_A(V, N, Call, ...) \
  if ((V = Call(__VA_ARGS__)) == N) { perror(#Call); exit(1); }
#define SYS_dA(V, Call, ...) SYS_A(V, -1, Call, __VA_ARGS__)
#define SYS_d(Call, ...) \
  if (Call(__VA_ARGS__) == -1) { perror(#Call); exit(1); }

void write_pgm(char *filename, greyimage img, unsigned w, unsigned h) {
    int fd, len; char buf[64];

    len = snprintf(buf, 64, "P5\n%u %u\n255\n", w, h);
    SYS_dA(fd, open, filename, O_WRONLY|O_TRUNC|O_CREAT, 0666);
    SYS_d(write, fd, buf, len);
    SYS_d(write, fd, img, w*h);
    SYS_d(close, fd);
}

rgbimage read_ppm(char *filename, unsigned *w, unsigned *h) {
    int fd, sz;  char *p, *data; char buf[64];

    SYS_dA(fd, open, filename, O_RDONLY);
    SYS_d(read, fd, buf, 64);
    *w = strtoul(buf + 3, &p, 10); // width, skip header
    *h = strtoul(p + 1, &p, 10); // height
    (void)strtoul(p + 1, &p, 10); // skip depth
    sz = *w * *h * 3 + (p + 1 - buf);
    SYS_A(data, MAP_FAILED, mmap, 0, sz, PROT_READ, MAP_FILE|MAP_PRIVATE, fd, 0);
    SYS_d(close, fd);
    return (rgbimage)(data + (p + 1 - buf));
}
\end{lstlisting}

\begin{lstlisting}[float,label=lst:gp,caption={plot.gp: GNUplot script used to
  generate the diagrams in this article.},language=gnuplot,escapechar=?,basicstyle=\tiny]
set term pdf
set logscale x
set logscale y
set ytics format "%.1e"
set title "Performance\nRunning on MacBookPro7,1 :: OSX 10.6.8 :: Intel Core 2 Duo @ 2.4GHz :: 3MB cache"
set xlabel "Image size (total number of pixels)"
set ylabel "Execution time (seconds)"

set output "nperf-hs.pdf"
plot "hs/summary" using 1:3 title "Haskell, GHC 7.4.2, DevIL I/O" 

set title "Normalized performance\nRunning on MacBookPro7,1 :: OSX 10.6.8 :: Intel Core 2 Duo @ 2.4GHz :: 3MB cache"
set ylabel "Seconds per pixel converted"
set output "nperf-hs2.pdf"
plot "hs/summary" using 1:($3/$1) title "Haskell, GHC 7.4.2, DevIL I/O" 

set output "nperf-lang.pdf"
plot "hs/summary" using 1:($3/$1) title "Haskell, GHC 7.4.2, DevIL I/O", \
     "c-clang-o3-il-double2d/summary" using 1:($3/$1) title "C, DP, RM 2D, Clang 3.3, -O3, DevIL I/O", \
     "c-clang-o3-unix-double2d/summary" using 1:($3/$1) title "C, DP, RM 2D, Clang 3.3, -O3, mmap/write"

set output "nperf-empty.pdf"
plot "c-clang-o3-il-double2d/summary" using 1:($3/$1) title "C, DP, RM 2D, Clang 3.3, -O3, DevIL I/O", \
     "c-clang-o3-unix-double2d/summary" using 1:($3/$1) title "C, DP, RM 2D, Clang 3.3, -O3, mmap/write", \
     "c-clang-o3-il-empty/summary" using 1:($3/$1) title "C, DP, RM 2D, Clang 3.3, -O3, DevIL I/O, no compute", \
     "c-clang-o3-unix-empty/summary" using 1:($3/$1) title "C, DP, RM 2D, Clang 3.3, -O3, mmap/write, no compute"

set yrange [*:1e-6]
set output "nperf-trans.pdf"
plot "c-clang-o3-unix-double2d/summary" using 1:($3/$1) title "C, DP, RM 2D, Clang 3.3, -O3, mmap/write", \
     "c-clang-o3-unix-double2dt/summary" using 1:($3/$1) title "C, DP, CM 2D, Clang 3.3, -O3, mmap/write"

set yrange [*:1e-7]
set output "nperf-dim.pdf"
plot "c-clang-o3-unix-double2d/summary" using 1:($3/$1) title "C, DP, RM 2D, Clang 3.3, -O3, mmap/write", \
     "c-clang-o3-unix-double1d/summary" using 1:($3/$1) title "C, DP, 1D, Clang 3.3, -O3, mmap/write"

set output "nperf-type.pdf"
plot "c-clang-o3-unix-double1d/summary" using 1:($3/$1) title "C, DP, 1D, Clang 3.3, -O3, mmap/write", \
     "c-clang-o3-unix-float1d/summary" using 1:($3/$1) title "C, SP, 1D, Clang 3.3, -O3, mmap/write", \
     "c-clang-o3-unix-int1d/summary" using 1:($3/$1) title "C, int, 1D, Clang 3.3, -O3, mmap/write"

set output "nperf-comp.pdf"
plot "c-clang-o3-unix-int1d/summary" using 1:($3/$1) title "C, int, 1D, Clang 3.3, -O3, mmap/write", \
     "c-gcc-o3-unix-int1d/summary" using 1:($3/$1) title "C, int, 1D, GCC 4.8, -O3, mmap/write",  \
     "c-clang-o3-unix-double1d/summary" using 1:($3/$1) title "C, DP, 1D, Clang 3.3, -O3, mmap/write", \
     "c-gcc-o3-unix-double1d/summary" using 1:($3/$1) title "C, DP, 1D, GCC 4.8, -O3, mmap/write"

set output "nperf-flags.pdf"
plot "c-clang-base-unix-int1d/summary" using 1:($3/$1) title "C, int, 1D, Clang 3.3, no opts, mmap/write", \
     "c-clang-o1-unix-int1d/summary" using 1:($3/$1) title "C, int, 1D, Clang 3.3, -O1, mmap/write", \
     "c-clang-o3-unix-int1d/summary" using 1:($3/$1) title "C, int, 1D, Clang 3.3, -O3, mmap/write"

set output "nperf-align.pdf"
plot "c-clang-o3-unix-double2d/summary" using 1:($2/$1) title "C, int, 1D, Clang 3.3, -O3, mmap/write", \
     "c-clang-o3-unix2-double2d/summary" using 1:($2/$1) title "C, int, 1D, Clang 3.3, -O3, malloc+read/write"

unset logscale y
set yrange [0:1e-07]
set output "nperf-sub.pdf"
plot  "c-clang-o3-unix-double2d/summary" using 1:($2/$1) title "C, DP, RM 2D, Clang 3.3, -O3, mmap/write (I/O+compute)", \
      "c-clang-o3-unix-empty/summary" using 1:($2/$1) title "C, DP, RM 2D, Clang 3.3, -O3, mmap/write, no compute (I/O only)", \
      "< join c-clang-o3-unix-double2d/summary c-clang-o3-unix-empty/summary" \
         using 1:(($2-$7)/$1) title "C, DP, RM 2D, Clang 3.3, -O3, I/O substracted (computation only)"

set logscale y
set yrange [*:5e-6]
set output "nperf-sub2.pdf"
plot  "c-clang-o3-il-double2d/summary" using 1:($3/$1) title "C, DP, RM 2D, Clang 3.3, -O3, DevIL (I/O+compute)", \
     "c-clang-o3-il-empty/summary" using 1:($3/$1) title "C, DP, RM 2D, Clang 3.3, -O3, DevIL, no compute (I/O only)", \
     "< join c-clang-o3-il-double2d/summary c-clang-o3-il-empty/summary" \
        using 1:(($3-$8)/$1) title "C, DP, RM 2D, Clang 3.3, -O3, I/O substracted (computation only)"

unset logscale y
set title "Normalized performance (I/O removed)\nRunning on MacBookPro7,1 :: OSX 10.6.8 :: Intel Core 2 Duo @ 2.4GHz :: 3MB cache"
set yrange [0:1e-07]
set output "nperf-sub-trans.pdf"
plot "< join c-clang-o3-unix-double2d/summary c-clang-o3-unix-empty/summary"  \
           using 1:(($3-$8)/$1) title "C, DP, RM 2D, Clang 3.3, -O3, I/O substracted (computation only)", \
     "< join c-clang-o3-unix-double2dt/summary c-clang-o3-unix-empty/summary"  \
           using 1:(($3-$8)/$1) title "C, DP, CM 2D, Clang 3.3, -O3, I/O substracted (computation only)"

set yrange [0:4e-8]
set output "nperf-sub-dim.pdf"
plot "< join c-clang-o3-unix-double2d/summary c-clang-o3-unix-empty/summary"  \
         using 1:(($3-$8)/$1) title "C, DP, RM 2D, Clang 3.3, -O3, I/O substracted (computation only)", \
     "< join c-clang-o3-unix-double1d/summary c-clang-o3-unix-empty/summary"  \
        using 1:(($3-$8)/$1) title "C, DP, 1D, Clang 3.3, -O3, I/O substracted (computation only)"

set yrange [0:2e-8]
set output "nperf-sub-type.pdf"
plot "< join c-clang-o3-unix-double1d/summary c-clang-o3-unix-empty/summary" \
        using 1:(($3-$8)/$1) title "C, DP, RM 2D, Clang 3.3, -O3, I/O substracted (computation only)", \
     "< join c-clang-o3-unix-float1d/summary c-clang-o3-unix-empty/summary" \
        using 1:(($3-$8)/$1) title "C, SP, RM 2D, Clang 3.3, -O3, I/O substracted (computation only)", \
     "< join c-clang-o3-unix-int1d/summary c-clang-o3-unix-empty/summary" \
        using 1:(($3-$8)/$1) title "C, int, 1D, Clang 3.3, -O3, I/O substracted (computation only)"

set yrange [0:2e-8]
set output "nperf-sub-flags.pdf"
plot "< join c-clang-base-unix-int1d/summary c-clang-base-unix-empty/summary" \
         using 1:(($3-$8)/$1) title "C, int, 1D, Clang 3.3, no opts, I/O substracted (computation only)", \
     "< join c-clang-o1-unix-int1d/summary c-clang-o1-unix-empty/summary" \
        using 1:(($3-$8)/$1) title "C, int, 1D, Clang 3.3, -O1, I/O substracted (computation only)", \
     "< join c-clang-o3-unix-int1d/summary c-clang-o3-unix-empty/summary" \
        using 1:(($3-$8)/$1) title "C, int, 1D, Clang 3.3, -O3, I/O substracted (computation only)"
     
\end{lstlisting}

\begin{lstlisting}[float,label=lst:gp,caption={GNUmakefile used to
  generate the data for the diagrams.},language=make,escapechar=?,basicstyle=\tiny]
GHC = ghc
CC_gcc = gcc-mp-4.8
CC_clang = clang-mp-3.3
CFLAGS_base =
CFLAGS_o1 = -O1
CFLAGS_o3 = -O3
CFLAGS_fast = -O3 -fomit-frame-pointer -ftree-vectorize -ffast-math -fstrict-aliasing
LD_unix =
LD_il = -lil -L/opt/local/lib -I/opt/local/include
DOTIME = ./mytime

PPMDIR = $(HOME)/src/im
PPMS := $(wildcard $(PPMDIR)/*.ppm)
SZS := $(shell cat sz)

v_at = $(v_at_$(V))
v_at_ = $(v_at_0)
v_at_0 = @
v_gen = $(v_gen_$(V))
v_gen_ = $(v_gen_0)
v_gen_0 = @echo '  GEN  $@';

ALGS = double2d
COMPILERS = gcc clang
FLAGLISTS = base o1 o3 fast
PIXS := $(patsubst ppm_%.c,%,$(wildcard ppm_*.c))

TDIRS := $(shell cat plot.gp|tr ' ' '\n'|grep '/summary'|tr -d '\"'|cut -d/ -f1|sort -u)

TARGETS = mytime $(foreach T,$(TDIRS), \
              $(T)/bench $(T)/summary \
	      $(foreach S,$(SZS), \
                  $(T)/$(S).res \
		  $(T)/$(S).times)) tables.tex

.PRECIOUS: bench %.ppm %.out %.res %.times summary

.SUFFIXES: .ppm .c .out .hs .res .times

all: $(TARGETS)

hs/bench: bench.hs
	$(v_at)mkdir -p hs
	$(v_gen)$(GHC) bench.hs -o $@

%.ppm:
	$(v_gen)convert -size $* plasma:fractal -blur 0x10 -emboss 2 -compress none $@

%.times: %.res
	$(v_gen)python -c "l=[`cat $^ | grep '^real' | cut -c5- | tr '\n' ','`]; s='$(notdir $*)'.split('x'); \
		print int(s[0])*int(s[1]),float(sum(l))/len(l),min(l),max(l),s[0],s[1]" >$@.tmp
	$(v_at)mv -f $@.tmp $@

.SECONDEXPANSION:

c-%/bench: bench.c ppm_$$(word 3,$$(subst -, ,$$*)).c $$(word 4,$$(subst -, ,$$*)).c
	$(v_at)mkdir -p c-$*
	$(v_gen)cd c-$*/ && \
	  ln -sf ../bench.c && \
	  ln -sf ../ppm_$(word 3,$(subst -, ,$*)).c ppm.c && \
	  ln -sf ../$(word 4,$(subst -, ,$*)).c compute.c && \
	   $(CC_$(word 1,$(subst -, ,$*))) \
	      $(CFLAGS_$(word 2,$(subst -, ,$*))) \
	      -o bench bench.c \
	      $(LD_$(word 3,$(subst -, ,$*)))

%.res: $$(foreach I,0 1 2 3 4,$$(dir $$@)$$*.$$(I).out)
	$(v_gen)cat $^ >$@.tmp
	$(v_at)mv $@.tmp $@

%.out: $$(dir $$@)bench $(PPMDIR)/$$(word 1,$$(subst ., ,$$(notdir $$*))).ppm
	$(v_gen)$(DOTIME) $^ /tmp/$(notdir $@).pgm 2>$@.tmp
	$(v_at)mv -f $@.tmp $@
	$(v_at)rm -f /tmp/$(notdir $@).pgm

%/summary: $$(foreach S,$(SZS),$$*/$$(S).times)
	$(v_gen)cat $^ >$@.tmp
	$(v_at)mv $@.tmp $@

%/table: %/summary
	$(v_gen)sed -e 's/ / \& /g;s/$$/ \\\\/g' <$^ >$@.tmp
	$(v_at)mv $@.tmp $@

tables.tex: $(TDIRS:%=%/table)
	$(v_gen)for t in $(TDIRS); do \
	     printf '\\begin{table}\n\\centering\n\\begin{tabular}{llllll}\n'; \
	     printf 'Pixels&Avg&Min&Max&Width&Height \\\\\n\\hline\n'; \
	     cat $$t/table; \
	     printf '\\end{tabular}\n\\caption{%s}\\label{tab:res:%s}\n' $$t/summary $$t/summary; \
	     printf '\\end{table}\n\\clearpage\n'; \
	done >$@.tmp
	$(v_at)mv $@.tmp $@
\end{lstlisting}

\begin{table}
\centering
\begin{tabular}{llllll}
Pixels&Avg&Min&Max&Width&Height \\
\hline
10 & 0.0013862 & 0.001295 & 0.001686 & 5 & 2 \\
16 & 0.001486 & 0.001422 & 0.001567 & 4 & 4 \\
20 & 0.0013946 & 0.00135 & 0.001471 & 5 & 4 \\
32 & 0.001415 & 0.001323 & 0.001478 & 8 & 4 \\
48 & 0.0014936 & 0.001336 & 0.001598 & 8 & 6 \\
64 & 0.0015456 & 0.001468 & 0.001724 & 8 & 8 \\
96 & 0.0014484 & 0.001368 & 0.001585 & 16 & 6 \\
128 & 0.001585 & 0.00147 & 0.001697 & 16 & 8 \\
192 & 0.0016648 & 0.001458 & 0.001819 & 16 & 12 \\
256 & 0.00165 & 0.001545 & 0.001795 & 16 & 16 \\
480 & 0.001527 & 0.001445 & 0.001631 & 32 & 15 \\
512 & 0.001443 & 0.001376 & 0.001493 & 32 & 16 \\
992 & 0.001426 & 0.001342 & 0.001517 & 32 & 31 \\
1024 & 0.0015804 & 0.001384 & 0.002083 & 32 & 32 \\
1984 & 0.001486 & 0.001355 & 0.001616 & 64 & 31 \\
2048 & 0.0014586 & 0.001436 & 0.001469 & 64 & 32 \\
4096 & 0.0014974 & 0.00148 & 0.001517 & 64 & 64 \\
4992 & 0.0015038 & 0.001431 & 0.001665 & 128 & 39 \\
8192 & 0.0015616 & 0.001524 & 0.001594 & 128 & 64 \\
9984 & 0.0014896 & 0.001386 & 0.001653 & 128 & 78 \\
16384 & 0.0015948 & 0.001502 & 0.001717 & 128 & 128 \\
19968 & 0.001504 & 0.001442 & 0.00164 & 256 & 78 \\
32768 & 0.0015162 & 0.001484 & 0.001582 & 256 & 128 \\
49920 & 0.0016716 & 0.001547 & 0.001765 & 256 & 195 \\
65536 & 0.0015904 & 0.001573 & 0.001604 & 256 & 256 \\
99840 & 0.0018216 & 0.001724 & 0.001972 & 512 & 195 \\
131072 & 0.001965 & 0.00184 & 0.002151 & 512 & 256 \\
199680 & 0.0023032 & 0.00207 & 0.002633 & 512 & 390 \\
262144 & 0.002405 & 0.002215 & 0.002688 & 512 & 512 \\
499712 & 0.0046076 & 0.002726 & 0.005595 & 1024 & 488 \\
524288 & 0.0115674 & 0.006185 & 0.018067 & 1024 & 512 \\
999424 & 0.021863 & 0.02042 & 0.022241 & 1024 & 976 \\
1048576 & 0.0222758 & 0.022218 & 0.022323 & 1024 & 1024 \\
1998848 & 0.0260388 & 0.025325 & 0.026787 & 2048 & 976 \\
2097152 & 0.0266788 & 0.023063 & 0.027721 & 2048 & 1024 \\
4194304 & 0.0366048 & 0.035857 & 0.037427 & 2048 & 2048 \\
4999168 & 0.0415548 & 0.040989 & 0.042722 & 2048 & 2441 \\
8388608 & 0.0369938 & 0.036359 & 0.037304 & 4096 & 2048 \\
9998336 & 0.0428376 & 0.041531 & 0.045774 & 4096 & 2441 \\
16777216 & 0.1026934 & 0.100651 & 0.105384 & 4096 & 4096 \\
\end{tabular}
\caption{c-clang-base-unix-empty/summary}\label{tab:res:c-clang-base-unix-empty/summary}
\end{table}
\clearpage
\begin{table}
\centering
\begin{tabular}{llllll}
Pixels&Avg&Min&Max&Width&Height \\
\hline
10 & 0.0014302 & 0.001364 & 0.001482 & 5 & 2 \\
16 & 0.0015982 & 0.001492 & 0.001683 & 4 & 4 \\
20 & 0.0014596 & 0.001338 & 0.001626 & 5 & 4 \\
32 & 0.0015564 & 0.001382 & 0.001733 & 8 & 4 \\
48 & 0.0015978 & 0.001533 & 0.001679 & 8 & 6 \\
64 & 0.0015974 & 0.001498 & 0.00176 & 8 & 8 \\
96 & 0.0014994 & 0.001389 & 0.001598 & 16 & 6 \\
128 & 0.0015662 & 0.00144 & 0.001682 & 16 & 8 \\
192 & 0.0015644 & 0.001494 & 0.001647 & 16 & 12 \\
256 & 0.0014034 & 0.001337 & 0.001465 & 16 & 16 \\
480 & 0.0016188 & 0.001492 & 0.001748 & 32 & 15 \\
512 & 0.0014758 & 0.001333 & 0.001618 & 32 & 16 \\
992 & 0.001501 & 0.001408 & 0.001673 & 32 & 31 \\
1024 & 0.0016836 & 0.001578 & 0.001864 & 32 & 32 \\
1984 & 0.0015618 & 0.00147 & 0.001678 & 64 & 31 \\
2048 & 0.0016128 & 0.00146 & 0.00187 & 64 & 32 \\
4096 & 0.0015764 & 0.001496 & 0.001736 & 64 & 64 \\
4992 & 0.0016472 & 0.001548 & 0.001739 & 128 & 39 \\
8192 & 0.001651 & 0.001464 & 0.001733 & 128 & 64 \\
9984 & 0.0016694 & 0.001527 & 0.001824 & 128 & 78 \\
16384 & 0.0018418 & 0.001683 & 0.001956 & 128 & 128 \\
19968 & 0.0018998 & 0.001811 & 0.001995 & 256 & 78 \\
32768 & 0.0020082 & 0.001869 & 0.002137 & 256 & 128 \\
49920 & 0.0021614 & 0.002056 & 0.002338 & 256 & 195 \\
65536 & 0.002467 & 0.002236 & 0.002948 & 256 & 256 \\
99840 & 0.002718 & 0.002616 & 0.002932 & 512 & 195 \\
131072 & 0.0031828 & 0.002895 & 0.003412 & 512 & 256 \\
199680 & 0.0044898 & 0.004144 & 0.004967 & 512 & 390 \\
262144 & 0.005367 & 0.004872 & 0.005848 & 512 & 512 \\
499712 & 0.008919 & 0.007325 & 0.010794 & 1024 & 488 \\
524288 & 0.0222584 & 0.013031 & 0.025174 & 1024 & 512 \\
999424 & 0.0323036 & 0.031091 & 0.033689 & 1024 & 976 \\
1048576 & 0.0330452 & 0.032539 & 0.033536 & 1024 & 1024 \\
1998848 & 0.0463276 & 0.046022 & 0.046634 & 2048 & 976 \\
2097152 & 0.0479576 & 0.047376 & 0.048532 & 2048 & 1024 \\
4194304 & 0.0754948 & 0.066893 & 0.079124 & 2048 & 2048 \\
4999168 & 0.0869756 & 0.084773 & 0.089423 & 2048 & 2441 \\
8388608 & 0.1318272 & 0.129927 & 0.133747 & 4096 & 2048 \\
9998336 & 0.156257 & 0.153618 & 0.159581 & 4096 & 2441 \\
16777216 & 0.256308 & 0.252784 & 0.261073 & 4096 & 4096 \\
\end{tabular}
\caption{c-clang-base-unix-int1d/summary}\label{tab:res:c-clang-base-unix-int1d/summary}
\end{table}
\clearpage
\begin{table}
\centering
\begin{tabular}{llllll}
Pixels&Avg&Min&Max&Width&Height \\
\hline
10 & 0.0014594 & 0.00139 & 0.00161 & 5 & 2 \\
16 & 0.001498 & 0.001445 & 0.001563 & 4 & 4 \\
20 & 0.0014568 & 0.001416 & 0.001491 & 5 & 4 \\
32 & 0.0014814 & 0.00136 & 0.001577 & 8 & 4 \\
48 & 0.0013826 & 0.001344 & 0.001411 & 8 & 6 \\
64 & 0.0014394 & 0.001353 & 0.001563 & 8 & 8 \\
96 & 0.0013904 & 0.001333 & 0.001462 & 16 & 6 \\
128 & 0.0014576 & 0.00141 & 0.001503 & 16 & 8 \\
192 & 0.0014982 & 0.001413 & 0.001587 & 16 & 12 \\
256 & 0.001431 & 0.001347 & 0.001486 & 16 & 16 \\
480 & 0.0014872 & 0.001408 & 0.001526 & 32 & 15 \\
512 & 0.0013802 & 0.001325 & 0.001449 & 32 & 16 \\
992 & 0.0014756 & 0.001441 & 0.001513 & 32 & 31 \\
1024 & 0.001469 & 0.001392 & 0.001527 & 32 & 32 \\
1984 & 0.001428 & 0.001348 & 0.001491 & 64 & 31 \\
2048 & 0.0014496 & 0.001378 & 0.001513 & 64 & 32 \\
4096 & 0.0015136 & 0.001452 & 0.001568 & 64 & 64 \\
4992 & 0.001461 & 0.001418 & 0.001508 & 128 & 39 \\
8192 & 0.0015428 & 0.001433 & 0.001605 & 128 & 64 \\
9984 & 0.0014892 & 0.001415 & 0.001538 & 128 & 78 \\
16384 & 0.001661 & 0.001568 & 0.001737 & 128 & 128 \\
19968 & 0.0015614 & 0.001486 & 0.001687 & 256 & 78 \\
32768 & 0.0017076 & 0.001511 & 0.001905 & 256 & 128 \\
49920 & 0.0016304 & 0.001613 & 0.001646 & 256 & 195 \\
65536 & 0.0016016 & 0.001569 & 0.001646 & 256 & 256 \\
99840 & 0.0017068 & 0.001659 & 0.001779 & 512 & 195 \\
131072 & 0.0018224 & 0.001741 & 0.00186 & 512 & 256 \\
199680 & 0.0023806 & 0.002064 & 0.002592 & 512 & 390 \\
262144 & 0.0025998 & 0.002176 & 0.002904 & 512 & 512 \\
499712 & 0.0036716 & 0.00288 & 0.004395 & 1024 & 488 \\
524288 & 0.0098236 & 0.004753 & 0.016192 & 1024 & 512 \\
999424 & 0.022286 & 0.022051 & 0.022402 & 1024 & 976 \\
1048576 & 0.022048 & 0.020388 & 0.022666 & 1024 & 1024 \\
1998848 & 0.0259494 & 0.023757 & 0.027333 & 2048 & 976 \\
2097152 & 0.0277118 & 0.027503 & 0.027954 & 2048 & 1024 \\
4194304 & 0.0349692 & 0.030343 & 0.036794 & 2048 & 2048 \\
4999168 & 0.040798 & 0.038609 & 0.041924 & 2048 & 2441 \\
8388608 & 0.0358476 & 0.035465 & 0.036247 & 4096 & 2048 \\
9998336 & 0.0390578 & 0.038351 & 0.040156 & 4096 & 2441 \\
16777216 & 0.1004596 & 0.09854 & 0.103355 & 4096 & 4096 \\
\end{tabular}
\caption{c-clang-o1-unix-empty/summary}\label{tab:res:c-clang-o1-unix-empty/summary}
\end{table}
\clearpage
\begin{table}
\centering
\begin{tabular}{llllll}
Pixels&Avg&Min&Max&Width&Height \\
\hline
10 & 0.0014624 & 0.001366 & 0.001593 & 5 & 2 \\
16 & 0.0016956 & 0.001512 & 0.002112 & 4 & 4 \\
20 & 0.0015232 & 0.001421 & 0.001605 & 5 & 4 \\
32 & 0.0014544 & 0.001388 & 0.001624 & 8 & 4 \\
48 & 0.0015852 & 0.001517 & 0.001652 & 8 & 6 \\
64 & 0.0014448 & 0.001357 & 0.001574 & 8 & 8 \\
96 & 0.0015802 & 0.001472 & 0.001697 & 16 & 6 \\
128 & 0.001517 & 0.00141 & 0.001599 & 16 & 8 \\
192 & 0.0014906 & 0.001407 & 0.001605 & 16 & 12 \\
256 & 0.0016178 & 0.001492 & 0.001736 & 16 & 16 \\
480 & 0.0015232 & 0.001334 & 0.00193 & 32 & 15 \\
512 & 0.0014658 & 0.001403 & 0.001598 & 32 & 16 \\
992 & 0.001555 & 0.001459 & 0.001673 & 32 & 31 \\
1024 & 0.0014768 & 0.001374 & 0.00163 & 32 & 32 \\
1984 & 0.0016586 & 0.001609 & 0.001747 & 64 & 31 \\
2048 & 0.001503 & 0.001367 & 0.00163 & 64 & 32 \\
4096 & 0.0016526 & 0.001559 & 0.00175 & 64 & 64 \\
4992 & 0.0015192 & 0.001459 & 0.001559 & 128 & 39 \\
8192 & 0.0017076 & 0.001573 & 0.001908 & 128 & 64 \\
9984 & 0.0016656 & 0.001596 & 0.001723 & 128 & 78 \\
16384 & 0.0016672 & 0.001546 & 0.001809 & 128 & 128 \\
19968 & 0.0018394 & 0.001712 & 0.002011 & 256 & 78 \\
32768 & 0.001784 & 0.001651 & 0.001932 & 256 & 128 \\
49920 & 0.0020034 & 0.001868 & 0.002125 & 256 & 195 \\
65536 & 0.002088 & 0.001985 & 0.002229 & 256 & 256 \\
99840 & 0.0023442 & 0.002209 & 0.002583 & 512 & 195 \\
131072 & 0.0024804 & 0.002251 & 0.00282 & 512 & 256 \\
199680 & 0.0031476 & 0.002907 & 0.003404 & 512 & 390 \\
262144 & 0.0037512 & 0.003289 & 0.004103 & 512 & 512 \\
499712 & 0.0060208 & 0.004916 & 0.00689 & 1024 & 488 \\
524288 & 0.0230936 & 0.021627 & 0.0242 & 1024 & 512 \\
999424 & 0.0286376 & 0.027709 & 0.029762 & 1024 & 976 \\
1048576 & 0.028526 & 0.028034 & 0.028999 & 1024 & 1024 \\
1998848 & 0.0389484 & 0.038404 & 0.039874 & 2048 & 976 \\
2097152 & 0.0405994 & 0.039878 & 0.041356 & 2048 & 1024 \\
4194304 & 0.061057 & 0.05849 & 0.063393 & 2048 & 2048 \\
4999168 & 0.066658 & 0.065538 & 0.067487 & 2048 & 2441 \\
8388608 & 0.0994106 & 0.094852 & 0.103462 & 4096 & 2048 \\
9998336 & 0.118118 & 0.116988 & 0.118876 & 4096 & 2441 \\
16777216 & 0.1819934 & 0.1781 & 0.184849 & 4096 & 4096 \\
\end{tabular}
\caption{c-clang-o1-unix-int1d/summary}\label{tab:res:c-clang-o1-unix-int1d/summary}
\end{table}
\clearpage
\begin{table}
\centering
\begin{tabular}{llllll}
Pixels&Avg&Min&Max&Width&Height \\
\hline
10 & 0.0030938 & 0.002669 & 0.003804 & 5 & 2 \\
16 & 0.0027886 & 0.002519 & 0.00292 & 4 & 4 \\
20 & 0.0028598 & 0.002593 & 0.003175 & 5 & 4 \\
32 & 0.0026324 & 0.002522 & 0.002769 & 8 & 4 \\
48 & 0.002688 & 0.002388 & 0.002895 & 8 & 6 \\
64 & 0.0028188 & 0.002722 & 0.002977 & 8 & 8 \\
96 & 0.002798 & 0.002628 & 0.003018 & 16 & 6 \\
128 & 0.002772 & 0.002652 & 0.002861 & 16 & 8 \\
192 & 0.0027476 & 0.00262 & 0.002886 & 16 & 12 \\
256 & 0.0027978 & 0.00271 & 0.002937 & 16 & 16 \\
480 & 0.003189 & 0.002842 & 0.004052 & 32 & 15 \\
512 & 0.0030602 & 0.002886 & 0.003373 & 32 & 16 \\
992 & 0.0033292 & 0.003159 & 0.003432 & 32 & 31 \\
1024 & 0.0030904 & 0.003007 & 0.003211 & 32 & 32 \\
1984 & 0.0038904 & 0.003713 & 0.004246 & 64 & 31 \\
2048 & 0.0038656 & 0.003643 & 0.003967 & 64 & 32 \\
4096 & 0.0050104 & 0.004768 & 0.005289 & 64 & 64 \\
4992 & 0.0055754 & 0.005388 & 0.005753 & 128 & 39 \\
8192 & 0.0074466 & 0.007105 & 0.008035 & 128 & 64 \\
9984 & 0.0085498 & 0.008445 & 0.008655 & 128 & 78 \\
16384 & 0.0118718 & 0.011601 & 0.012064 & 128 & 128 \\
19968 & 0.0140644 & 0.013803 & 0.014551 & 256 & 78 \\
32768 & 0.0208766 & 0.020533 & 0.021432 & 256 & 128 \\
49920 & 0.0302314 & 0.02995 & 0.030495 & 256 & 195 \\
65536 & 0.038837 & 0.038649 & 0.039379 & 256 & 256 \\
99840 & 0.0577934 & 0.057091 & 0.058998 & 512 & 195 \\
131072 & 0.074207 & 0.0737 & 0.075102 & 512 & 256 \\
199680 & 0.1273968 & 0.127007 & 0.128073 & 512 & 390 \\
262144 & 0.161762 & 0.161455 & 0.162801 & 512 & 512 \\
499712 & 0.2911702 & 0.29026 & 0.293007 & 1024 & 488 \\
524288 & 0.3044786 & 0.303479 & 0.305146 & 1024 & 512 \\
999424 & 0.5693138 & 0.567653 & 0.571057 & 1024 & 976 \\
1048576 & 0.5947214 & 0.59223 & 0.599434 & 1024 & 1024 \\
1998848 & 1.1224458 & 1.119447 & 1.126038 & 2048 & 976 \\
2097152 & 1.1723698 & 1.170106 & 1.175108 & 2048 & 1024 \\
4194304 & 2.345104 & 2.339499 & 2.351858 & 2048 & 2048 \\
4999168 & 2.7895248 & 2.784703 & 2.79531 & 2048 & 2441 \\
8388608 & 4.650901 & 4.642141 & 4.65494 & 4096 & 2048 \\
9998336 & 5.57949 & 5.534664 & 5.70067 & 4096 & 2441 \\
16777216 & 9.3506568 & 9.326829 & 9.368743 & 4096 & 4096 \\
\end{tabular}
\caption{c-clang-o3-il-double2d/summary}\label{tab:res:c-clang-o3-il-double2d/summary}
\end{table}
\clearpage
\begin{table}
\centering
\begin{tabular}{llllll}
Pixels&Avg&Min&Max&Width&Height \\
\hline
10 & 0.0028614 & 0.002416 & 0.003762 & 5 & 2 \\
16 & 0.002603 & 0.002395 & 0.002835 & 4 & 4 \\
20 & 0.0026944 & 0.002475 & 0.003013 & 5 & 4 \\
32 & 0.0027426 & 0.002532 & 0.002856 & 8 & 4 \\
48 & 0.0026356 & 0.002522 & 0.002878 & 8 & 6 \\
64 & 0.002729 & 0.002488 & 0.003075 & 8 & 8 \\
96 & 0.0028868 & 0.002587 & 0.003304 & 16 & 6 \\
128 & 0.002741 & 0.002462 & 0.003002 & 16 & 8 \\
192 & 0.0027296 & 0.002625 & 0.002856 & 16 & 12 \\
256 & 0.0028658 & 0.002682 & 0.003229 & 16 & 16 \\
480 & 0.0029024 & 0.002735 & 0.003042 & 32 & 15 \\
512 & 0.0029884 & 0.002833 & 0.003213 & 32 & 16 \\
992 & 0.0034404 & 0.003053 & 0.003858 & 32 & 31 \\
1024 & 0.0032878 & 0.003112 & 0.003628 & 32 & 32 \\
1984 & 0.0038852 & 0.003695 & 0.004162 & 64 & 31 \\
2048 & 0.0039464 & 0.003746 & 0.004253 & 64 & 32 \\
4096 & 0.0051052 & 0.004794 & 0.005309 & 64 & 64 \\
4992 & 0.0055894 & 0.005419 & 0.005763 & 128 & 39 \\
8192 & 0.0073938 & 0.007169 & 0.007728 & 128 & 64 \\
9984 & 0.0083842 & 0.008119 & 0.008803 & 128 & 78 \\
16384 & 0.0119462 & 0.011582 & 0.012285 & 128 & 128 \\
19968 & 0.013829 & 0.013699 & 0.013976 & 256 & 78 \\
32768 & 0.0207446 & 0.020411 & 0.021224 & 256 & 128 \\
49920 & 0.0299368 & 0.029384 & 0.030453 & 256 & 195 \\
65536 & 0.0384048 & 0.038327 & 0.038543 & 256 & 256 \\
99840 & 0.0570258 & 0.056526 & 0.057456 & 512 & 195 \\
131072 & 0.0735488 & 0.073271 & 0.074016 & 512 & 256 \\
199680 & 0.126947 & 0.126547 & 0.127116 & 512 & 390 \\
262144 & 0.1606156 & 0.1602 & 0.160871 & 512 & 512 \\
499712 & 0.2893616 & 0.288918 & 0.290025 & 1024 & 488 \\
524288 & 0.302117 & 0.301276 & 0.302866 & 1024 & 512 \\
999424 & 0.5622004 & 0.560459 & 0.563203 & 1024 & 976 \\
1048576 & 0.5891098 & 0.58682 & 0.591316 & 1024 & 1024 \\
1998848 & 1.1134272 & 1.106419 & 1.126347 & 2048 & 976 \\
2097152 & 1.1606344 & 1.156912 & 1.164354 & 2048 & 1024 \\
4194304 & 2.3315858 & 2.328636 & 2.336959 & 2048 & 2048 \\
4999168 & 2.7771312 & 2.771907 & 2.781743 & 2048 & 2441 \\
8388608 & 4.6288906 & 4.624046 & 4.633826 & 4096 & 2048 \\
9998336 & 5.506498 & 5.494406 & 5.51931 & 4096 & 2441 \\
16777216 & 9.3742154 & 9.263583 & 9.732905 & 4096 & 4096 \\
\end{tabular}
\caption{c-clang-o3-il-empty/summary}\label{tab:res:c-clang-o3-il-empty/summary}
\end{table}
\clearpage
\begin{table}
\centering
\begin{tabular}{llllll}
Pixels&Avg&Min&Max&Width&Height \\
\hline
10 & 0.0015176 & 0.001411 & 0.001592 & 5 & 2 \\
16 & 0.0014936 & 0.001359 & 0.001604 & 4 & 4 \\
20 & 0.0014908 & 0.001362 & 0.001679 & 5 & 4 \\
32 & 0.0015934 & 0.001493 & 0.001767 & 8 & 4 \\
48 & 0.0015948 & 0.001465 & 0.001732 & 8 & 6 \\
64 & 0.0015158 & 0.001444 & 0.001595 & 8 & 8 \\
96 & 0.0015378 & 0.001469 & 0.001581 & 16 & 6 \\
128 & 0.001646 & 0.001555 & 0.001744 & 16 & 8 \\
192 & 0.0016568 & 0.001563 & 0.001732 & 16 & 12 \\
256 & 0.0015218 & 0.001362 & 0.001679 & 16 & 16 \\
480 & 0.0015102 & 0.001436 & 0.001661 & 32 & 15 \\
512 & 0.0016384 & 0.00153 & 0.001722 & 32 & 16 \\
992 & 0.0015222 & 0.001434 & 0.001687 & 32 & 31 \\
1024 & 0.0016262 & 0.001473 & 0.001843 & 32 & 32 \\
1984 & 0.0015304 & 0.001403 & 0.001676 & 64 & 31 \\
2048 & 0.0014772 & 0.001387 & 0.001599 & 64 & 32 \\
4096 & 0.0014958 & 0.001414 & 0.001592 & 64 & 64 \\
4992 & 0.0016084 & 0.001545 & 0.001667 & 128 & 39 \\
8192 & 0.001666 & 0.001584 & 0.001767 & 128 & 64 \\
9984 & 0.0015942 & 0.001451 & 0.001666 & 128 & 78 \\
16384 & 0.0018006 & 0.001709 & 0.001874 & 128 & 128 \\
19968 & 0.0017904 & 0.001642 & 0.001888 & 256 & 78 \\
32768 & 0.0019014 & 0.001663 & 0.00215 & 256 & 128 \\
49920 & 0.0020784 & 0.001921 & 0.002195 & 256 & 195 \\
65536 & 0.0021692 & 0.001955 & 0.00238 & 256 & 256 \\
99840 & 0.0025544 & 0.002436 & 0.002874 & 512 & 195 \\
131072 & 0.0026178 & 0.002506 & 0.002818 & 512 & 256 \\
199680 & 0.0036118 & 0.003082 & 0.003889 & 512 & 390 \\
262144 & 0.004152 & 0.003791 & 0.004455 & 512 & 512 \\
499712 & 0.0068146 & 0.005608 & 0.008474 & 1024 & 488 \\
524288 & 0.0228782 & 0.022176 & 0.023958 & 1024 & 512 \\
999424 & 0.0281636 & 0.026574 & 0.029567 & 1024 & 976 \\
1048576 & 0.0298138 & 0.028936 & 0.030408 & 1024 & 1024 \\
1998848 & 0.0404544 & 0.038182 & 0.042113 & 2048 & 976 \\
2097152 & 0.042433 & 0.041778 & 0.042925 & 2048 & 1024 \\
4194304 & 0.063698 & 0.062337 & 0.064691 & 2048 & 2048 \\
4999168 & 0.0724326 & 0.071054 & 0.073824 & 2048 & 2441 \\
8388608 & 0.1083028 & 0.100587 & 0.112977 & 4096 & 2048 \\
9998336 & 0.1289222 & 0.127197 & 0.131232 & 4096 & 2441 \\
16777216 & 0.2078844 & 0.204441 & 0.212719 & 4096 & 4096 \\
\end{tabular}
\caption{c-clang-o3-unix-double1d/summary}\label{tab:res:c-clang-o3-unix-double1d/summary}
\end{table}
\clearpage
\begin{table}
\centering
\begin{tabular}{llllll}
Pixels&Avg&Min&Max&Width&Height \\
\hline
10 & 0.0014584 & 0.001363 & 0.0016 & 5 & 2 \\
16 & 0.0015908 & 0.001524 & 0.001654 & 4 & 4 \\
20 & 0.0015534 & 0.001472 & 0.00169 & 5 & 4 \\
32 & 0.001557 & 0.001497 & 0.001606 & 8 & 4 \\
48 & 0.0014506 & 0.001346 & 0.001546 & 8 & 6 \\
64 & 0.0015804 & 0.001413 & 0.001697 & 8 & 8 \\
96 & 0.0015498 & 0.001459 & 0.001613 & 16 & 6 \\
128 & 0.0015962 & 0.001473 & 0.001656 & 16 & 8 \\
192 & 0.0014516 & 0.001426 & 0.001507 & 16 & 12 \\
256 & 0.0015906 & 0.001493 & 0.001683 & 16 & 16 \\
480 & 0.0015446 & 0.001408 & 0.001671 & 32 & 15 \\
512 & 0.0016136 & 0.001578 & 0.001672 & 32 & 16 \\
992 & 0.001525 & 0.001451 & 0.001622 & 32 & 31 \\
1024 & 0.001593 & 0.001447 & 0.001861 & 32 & 32 \\
1984 & 0.0015972 & 0.001434 & 0.001698 & 64 & 31 \\
2048 & 0.001618 & 0.001447 & 0.001889 & 64 & 32 \\
4096 & 0.001692 & 0.001557 & 0.00189 & 64 & 64 \\
4992 & 0.0016606 & 0.001541 & 0.00172 & 128 & 39 \\
8192 & 0.0016338 & 0.001496 & 0.001818 & 128 & 64 \\
9984 & 0.001683 & 0.001611 & 0.001792 & 128 & 78 \\
16384 & 0.0017136 & 0.001577 & 0.001927 & 128 & 128 \\
19968 & 0.0018598 & 0.001668 & 0.002012 & 256 & 78 \\
32768 & 0.0019884 & 0.0018 & 0.002172 & 256 & 128 \\
49920 & 0.0021594 & 0.00208 & 0.002267 & 256 & 195 \\
65536 & 0.002199 & 0.002133 & 0.002397 & 256 & 256 \\
99840 & 0.0028378 & 0.002495 & 0.003422 & 512 & 195 \\
131072 & 0.002934 & 0.002811 & 0.003162 & 512 & 256 \\
199680 & 0.0039948 & 0.003582 & 0.004523 & 512 & 390 \\
262144 & 0.0048142 & 0.004471 & 0.005192 & 512 & 512 \\
499712 & 0.0080728 & 0.006251 & 0.010595 & 1024 & 488 \\
524288 & 0.0108182 & 0.007452 & 0.017111 & 1024 & 512 \\
999424 & 0.0301736 & 0.029703 & 0.03078 & 1024 & 976 \\
1048576 & 0.030701 & 0.030432 & 0.031062 & 1024 & 1024 \\
1998848 & 0.0419938 & 0.041877 & 0.042271 & 2048 & 976 \\
2097152 & 0.04396 & 0.04324 & 0.046318 & 2048 & 1024 \\
4194304 & 0.0700074 & 0.066876 & 0.074309 & 2048 & 2048 \\
4999168 & 0.078686 & 0.07379 & 0.08188 & 2048 & 2441 \\
8388608 & 0.1214904 & 0.118961 & 0.125023 & 4096 & 2048 \\
9998336 & 0.1415758 & 0.1324 & 0.150185 & 4096 & 2441 \\
16777216 & 0.235643 & 0.228462 & 0.246111 & 4096 & 4096 \\
\end{tabular}
\caption{c-clang-o3-unix-double2d/summary}\label{tab:res:c-clang-o3-unix-double2d/summary}
\end{table}
\clearpage
\begin{table}
\centering
\begin{tabular}{llllll}
Pixels&Avg&Min&Max&Width&Height \\
\hline
10 & 0.0015334 & 0.001346 & 0.00163 & 5 & 2 \\
16 & 0.0015126 & 0.001421 & 0.001597 & 4 & 4 \\
20 & 0.0016036 & 0.001532 & 0.001704 & 5 & 4 \\
32 & 0.0016424 & 0.001585 & 0.00181 & 8 & 4 \\
48 & 0.0015418 & 0.001492 & 0.001604 & 8 & 6 \\
64 & 0.0015788 & 0.001472 & 0.001697 & 8 & 8 \\
96 & 0.0016068 & 0.00151 & 0.001697 & 16 & 6 \\
128 & 0.0015532 & 0.001431 & 0.001687 & 16 & 8 \\
192 & 0.0016342 & 0.00145 & 0.001924 & 16 & 12 \\
256 & 0.001484 & 0.001405 & 0.001647 & 16 & 16 \\
480 & 0.0016652 & 0.001464 & 0.001881 & 32 & 15 \\
512 & 0.001569 & 0.001372 & 0.001666 & 32 & 16 \\
992 & 0.0015384 & 0.001417 & 0.001619 & 32 & 31 \\
1024 & 0.001481 & 0.001375 & 0.001726 & 32 & 32 \\
1984 & 0.0015272 & 0.001424 & 0.001626 & 64 & 31 \\
2048 & 0.0015894 & 0.001409 & 0.001673 & 64 & 32 \\
4096 & 0.001718 & 0.001604 & 0.00189 & 64 & 64 \\
4992 & 0.0018106 & 0.001641 & 0.00203 & 128 & 39 \\
8192 & 0.0018654 & 0.001711 & 0.002006 & 128 & 64 \\
9984 & 0.0016874 & 0.001604 & 0.001756 & 128 & 78 \\
16384 & 0.0018312 & 0.001579 & 0.002059 & 128 & 128 \\
19968 & 0.0019614 & 0.001772 & 0.002079 & 256 & 78 \\
32768 & 0.00218 & 0.002068 & 0.002342 & 256 & 128 \\
49920 & 0.0023692 & 0.002061 & 0.0026 & 256 & 195 \\
65536 & 0.0023206 & 0.002146 & 0.002473 & 256 & 256 \\
99840 & 0.0030044 & 0.002684 & 0.003467 & 512 & 195 \\
131072 & 0.0033824 & 0.003161 & 0.003641 & 512 & 256 \\
199680 & 0.0041384 & 0.003953 & 0.004286 & 512 & 390 \\
262144 & 0.0053512 & 0.005005 & 0.005899 & 512 & 512 \\
499712 & 0.0091248 & 0.007868 & 0.010416 & 1024 & 488 \\
524288 & 0.0097874 & 0.008554 & 0.011544 & 1024 & 512 \\
999424 & 0.0340112 & 0.032927 & 0.034901 & 1024 & 976 \\
1048576 & 0.035531 & 0.035199 & 0.03606 & 1024 & 1024 \\
1998848 & 0.0618162 & 0.058926 & 0.066473 & 2048 & 976 \\
2097152 & 0.0620528 & 0.058592 & 0.064776 & 2048 & 1024 \\
4194304 & 0.3670154 & 0.363387 & 0.37202 & 2048 & 2048 \\
4999168 & 0.463144 & 0.441961 & 0.482924 & 2048 & 2441 \\
8388608 & 0.8281012 & 0.791156 & 0.871089 & 4096 & 2048 \\
9998336 & 0.9910372 & 0.95868 & 1.037802 & 4096 & 2441 \\
16777216 & 1.6336186 & 1.601739 & 1.68 & 4096 & 4096 \\
\end{tabular}
\caption{c-clang-o3-unix-double2dt/summary}\label{tab:res:c-clang-o3-unix-double2dt/summary}
\end{table}
\clearpage
\begin{table}
\centering
\begin{tabular}{llllll}
Pixels&Avg&Min&Max&Width&Height \\
\hline
10 & 0.0014868 & 0.001328 & 0.001893 & 5 & 2 \\
16 & 0.0014554 & 0.001374 & 0.001562 & 4 & 4 \\
20 & 0.0013894 & 0.001336 & 0.001464 & 5 & 4 \\
32 & 0.0014236 & 0.00133 & 0.001457 & 8 & 4 \\
48 & 0.0014136 & 0.001323 & 0.001477 & 8 & 6 \\
64 & 0.0013378 & 0.001303 & 0.001363 & 8 & 8 \\
96 & 0.0014466 & 0.001402 & 0.001482 & 16 & 6 \\
128 & 0.0013288 & 0.001317 & 0.001345 & 16 & 8 \\
192 & 0.0014174 & 0.001349 & 0.001449 & 16 & 12 \\
256 & 0.0014294 & 0.001379 & 0.001493 & 16 & 16 \\
480 & 0.001443 & 0.001403 & 0.001518 & 32 & 15 \\
512 & 0.0014322 & 0.001349 & 0.001543 & 32 & 16 \\
992 & 0.0014506 & 0.001383 & 0.001491 & 32 & 31 \\
1024 & 0.0013734 & 0.001364 & 0.001389 & 32 & 32 \\
1984 & 0.0014948 & 0.001454 & 0.001549 & 64 & 31 \\
2048 & 0.001417 & 0.001352 & 0.001484 & 64 & 32 \\
4096 & 0.0015458 & 0.001388 & 0.001658 & 64 & 64 \\
4992 & 0.0015178 & 0.001406 & 0.001597 & 128 & 39 \\
8192 & 0.0014478 & 0.001381 & 0.001509 & 128 & 64 \\
9984 & 0.0015306 & 0.001408 & 0.001716 & 128 & 78 \\
16384 & 0.0015438 & 0.001474 & 0.001716 & 128 & 128 \\
19968 & 0.0015432 & 0.001462 & 0.001581 & 256 & 78 \\
32768 & 0.0015458 & 0.001476 & 0.001604 & 256 & 128 \\
49920 & 0.0016294 & 0.001574 & 0.001715 & 256 & 195 \\
65536 & 0.0016212 & 0.00156 & 0.001753 & 256 & 256 \\
99840 & 0.001713 & 0.001681 & 0.001754 & 512 & 195 \\
131072 & 0.0021076 & 0.001862 & 0.002332 & 512 & 256 \\
199680 & 0.0020978 & 0.002002 & 0.002321 & 512 & 390 \\
262144 & 0.0023818 & 0.002177 & 0.002688 & 512 & 512 \\
499712 & 0.002971 & 0.002727 & 0.003329 & 1024 & 488 \\
524288 & 0.0100456 & 0.004709 & 0.016563 & 1024 & 512 \\
999424 & 0.0220926 & 0.021561 & 0.022296 & 1024 & 976 \\
1048576 & 0.022294 & 0.022266 & 0.022358 & 1024 & 1024 \\
1998848 & 0.026321 & 0.024765 & 0.027103 & 2048 & 976 \\
2097152 & 0.0260974 & 0.022975 & 0.027501 & 2048 & 1024 \\
4194304 & 0.0345832 & 0.02967 & 0.036834 & 2048 & 2048 \\
4999168 & 0.0411714 & 0.041035 & 0.041264 & 2048 & 2441 \\
8388608 & 0.0361848 & 0.035027 & 0.036859 & 4096 & 2048 \\
9998336 & 0.0400812 & 0.039549 & 0.041217 & 4096 & 2441 \\
16777216 & 0.099768 & 0.098302 & 0.101399 & 4096 & 4096 \\
\end{tabular}
\caption{c-clang-o3-unix-empty/summary}\label{tab:res:c-clang-o3-unix-empty/summary}
\end{table}
\clearpage
\begin{table}
\centering
\begin{tabular}{llllll}
Pixels&Avg&Min&Max&Width&Height \\
\hline
10 & 0.0015476 & 0.001492 & 0.001612 & 5 & 2 \\
16 & 0.0016224 & 0.001563 & 0.001764 & 4 & 4 \\
20 & 0.001537 & 0.00146 & 0.001684 & 5 & 4 \\
32 & 0.0015836 & 0.001435 & 0.001721 & 8 & 4 \\
48 & 0.0015512 & 0.001431 & 0.001689 & 8 & 6 \\
64 & 0.0015934 & 0.001504 & 0.001696 & 8 & 8 \\
96 & 0.001535 & 0.001449 & 0.001648 & 16 & 6 \\
128 & 0.001587 & 0.001391 & 0.001845 & 16 & 8 \\
192 & 0.0015204 & 0.001386 & 0.001584 & 16 & 12 \\
256 & 0.0015664 & 0.00146 & 0.001692 & 16 & 16 \\
480 & 0.001514 & 0.001351 & 0.001785 & 32 & 15 \\
512 & 0.0015488 & 0.001428 & 0.001687 & 32 & 16 \\
992 & 0.0016128 & 0.001451 & 0.001789 & 32 & 31 \\
1024 & 0.0014926 & 0.001355 & 0.001671 & 32 & 32 \\
1984 & 0.0015744 & 0.001494 & 0.001656 & 64 & 31 \\
2048 & 0.0014874 & 0.001398 & 0.001572 & 64 & 32 \\
4096 & 0.001612 & 0.001486 & 0.001654 & 64 & 64 \\
4992 & 0.0015302 & 0.001423 & 0.001674 & 128 & 39 \\
8192 & 0.0016578 & 0.001517 & 0.001762 & 128 & 64 \\
9984 & 0.001631 & 0.001466 & 0.001803 & 128 & 78 \\
16384 & 0.0018496 & 0.001597 & 0.00199 & 128 & 128 \\
19968 & 0.0017378 & 0.001615 & 0.001841 & 256 & 78 \\
32768 & 0.0020268 & 0.001973 & 0.002074 & 256 & 128 \\
49920 & 0.0019714 & 0.001914 & 0.002012 & 256 & 195 \\
65536 & 0.002174 & 0.002012 & 0.002494 & 256 & 256 \\
99840 & 0.0025208 & 0.002296 & 0.002725 & 512 & 195 \\
131072 & 0.002626 & 0.002509 & 0.002897 & 512 & 256 \\
199680 & 0.0033054 & 0.003132 & 0.003635 & 512 & 390 \\
262144 & 0.003832 & 0.003615 & 0.004169 & 512 & 512 \\
499712 & 0.0071364 & 0.005794 & 0.007903 & 1024 & 488 \\
524288 & 0.0223346 & 0.015853 & 0.024698 & 1024 & 512 \\
999424 & 0.0281306 & 0.026406 & 0.029472 & 1024 & 976 \\
1048576 & 0.029095 & 0.028885 & 0.029281 & 1024 & 1024 \\
1998848 & 0.0398422 & 0.038995 & 0.040568 & 2048 & 976 \\
2097152 & 0.041057 & 0.040736 & 0.041471 & 2048 & 1024 \\
4194304 & 0.0621102 & 0.060263 & 0.064256 & 2048 & 2048 \\
4999168 & 0.069584 & 0.067825 & 0.071645 & 2048 & 2441 \\
8388608 & 0.1000746 & 0.087548 & 0.14594 & 4096 & 2048 \\
9998336 & 0.1221794 & 0.118968 & 0.125002 & 4096 & 2441 \\
16777216 & 0.1910922 & 0.186842 & 0.195253 & 4096 & 4096 \\
\end{tabular}
\caption{c-clang-o3-unix-float1d/summary}\label{tab:res:c-clang-o3-unix-float1d/summary}
\end{table}
\clearpage
\begin{table}
\centering
\begin{tabular}{llllll}
Pixels&Avg&Min&Max&Width&Height \\
\hline
10 & 0.0015284 & 0.001423 & 0.001707 & 5 & 2 \\
16 & 0.0015996 & 0.00139 & 0.001776 & 4 & 4 \\
20 & 0.0015838 & 0.001511 & 0.001626 & 5 & 4 \\
32 & 0.0015564 & 0.001399 & 0.001852 & 8 & 4 \\
48 & 0.0015952 & 0.001442 & 0.001748 & 8 & 6 \\
64 & 0.001532 & 0.001346 & 0.001782 & 8 & 8 \\
96 & 0.001573 & 0.00148 & 0.001678 & 16 & 6 \\
128 & 0.0015672 & 0.00143 & 0.001736 & 16 & 8 \\
192 & 0.0014888 & 0.001414 & 0.001661 & 16 & 12 \\
256 & 0.0014826 & 0.001432 & 0.001544 & 16 & 16 \\
480 & 0.0016216 & 0.001428 & 0.001869 & 32 & 15 \\
512 & 0.0014818 & 0.00141 & 0.00162 & 32 & 16 \\
992 & 0.0015578 & 0.001407 & 0.001667 & 32 & 31 \\
1024 & 0.0015774 & 0.001459 & 0.001657 & 32 & 32 \\
1984 & 0.0016122 & 0.001537 & 0.001735 & 64 & 31 \\
2048 & 0.0015406 & 0.001412 & 0.001597 & 64 & 32 \\
4096 & 0.0015542 & 0.001492 & 0.001635 & 64 & 64 \\
4992 & 0.001742 & 0.001644 & 0.001923 & 128 & 39 \\
8192 & 0.0015942 & 0.001505 & 0.001702 & 128 & 64 \\
9984 & 0.0016778 & 0.001541 & 0.001856 & 128 & 78 \\
16384 & 0.0017966 & 0.001739 & 0.001883 & 128 & 128 \\
19968 & 0.0018022 & 0.00158 & 0.001967 & 256 & 78 \\
32768 & 0.0018808 & 0.001741 & 0.002015 & 256 & 128 \\
49920 & 0.0020012 & 0.001725 & 0.002297 & 256 & 195 \\
65536 & 0.0021156 & 0.001866 & 0.002574 & 256 & 256 \\
99840 & 0.0025068 & 0.00223 & 0.002735 & 512 & 195 \\
131072 & 0.002692 & 0.002288 & 0.003329 & 512 & 256 \\
199680 & 0.00313 & 0.002748 & 0.003584 & 512 & 390 \\
262144 & 0.0038226 & 0.003381 & 0.004054 & 512 & 512 \\
499712 & 0.0072518 & 0.006375 & 0.008562 & 1024 & 488 \\
524288 & 0.022184 & 0.021806 & 0.022668 & 1024 & 512 \\
999424 & 0.0282942 & 0.027444 & 0.028935 & 1024 & 976 \\
1048576 & 0.027815 & 0.025846 & 0.02861 & 1024 & 1024 \\
1998848 & 0.0376148 & 0.035205 & 0.039317 & 2048 & 976 \\
2097152 & 0.039361 & 0.038343 & 0.04006 & 2048 & 1024 \\
4194304 & 0.0595208 & 0.058148 & 0.061289 & 2048 & 2048 \\
4999168 & 0.0655952 & 0.064573 & 0.067123 & 2048 & 2441 \\
8388608 & 0.1014334 & 0.099031 & 0.10332 & 4096 & 2048 \\
9998336 & 0.1155188 & 0.113595 & 0.117542 & 4096 & 2441 \\
16777216 & 0.18374 & 0.180268 & 0.186501 & 4096 & 4096 \\
\end{tabular}
\caption{c-clang-o3-unix-int1d/summary}\label{tab:res:c-clang-o3-unix-int1d/summary}
\end{table}
\clearpage
\begin{table}
\centering
\begin{tabular}{llllll}
Pixels&Avg&Min&Max&Width&Height \\
\hline
10 & 0.0015742 & 0.00153 & 0.001674 & 5 & 2 \\
16 & 0.0015042 & 0.001378 & 0.00168 & 4 & 4 \\
20 & 0.0016798 & 0.001619 & 0.001771 & 5 & 4 \\
32 & 0.001663 & 0.001472 & 0.001723 & 8 & 4 \\
48 & 0.0016186 & 0.001464 & 0.001764 & 8 & 6 \\
64 & 0.0016888 & 0.001561 & 0.001885 & 8 & 8 \\
96 & 0.0016822 & 0.001534 & 0.001883 & 16 & 6 \\
128 & 0.0015466 & 0.001356 & 0.001802 & 16 & 8 \\
192 & 0.001868 & 0.001633 & 0.002264 & 16 & 12 \\
256 & 0.0017654 & 0.00165 & 0.001855 & 16 & 16 \\
480 & 0.001756 & 0.001663 & 0.001867 & 32 & 15 \\
512 & 0.0018118 & 0.001706 & 0.001988 & 32 & 16 \\
992 & 0.0016124 & 0.001502 & 0.001752 & 32 & 31 \\
1024 & 0.0015718 & 0.001493 & 0.001649 & 32 & 32 \\
1984 & 0.001696 & 0.001521 & 0.001945 & 64 & 31 \\
2048 & 0.0016188 & 0.001455 & 0.001672 & 64 & 32 \\
4096 & 0.0018488 & 0.001591 & 0.002006 & 64 & 64 \\
4992 & 0.0018526 & 0.001771 & 0.001996 & 128 & 39 \\
8192 & 0.0018554 & 0.001693 & 0.001927 & 128 & 64 \\
9984 & 0.0018812 & 0.001656 & 0.002109 & 128 & 78 \\
16384 & 0.0018554 & 0.001672 & 0.002085 & 128 & 128 \\
19968 & 0.001788 & 0.001659 & 0.001926 & 256 & 78 \\
32768 & 0.00201 & 0.001874 & 0.002152 & 256 & 128 \\
49920 & 0.0023124 & 0.002142 & 0.002471 & 256 & 195 \\
65536 & 0.002615 & 0.002339 & 0.002801 & 256 & 256 \\
99840 & 0.0032086 & 0.00295 & 0.003498 & 512 & 195 \\
131072 & 0.0034318 & 0.003341 & 0.003574 & 512 & 256 \\
199680 & 0.004416 & 0.004213 & 0.004685 & 512 & 390 \\
262144 & 0.0055074 & 0.00523 & 0.005798 & 512 & 512 \\
499712 & 0.0097924 & 0.007824 & 0.011326 & 1024 & 488 \\
524288 & 0.0116496 & 0.008447 & 0.019813 & 1024 & 512 \\
999424 & 0.030933 & 0.030092 & 0.031646 & 1024 & 976 \\
1048576 & 0.0308166 & 0.030419 & 0.031295 & 1024 & 1024 \\
1998848 & 0.044111 & 0.042794 & 0.044722 & 2048 & 976 \\
2097152 & 0.0470242 & 0.046463 & 0.047269 & 2048 & 1024 \\
4194304 & 0.0764382 & 0.075429 & 0.077559 & 2048 & 2048 \\
4999168 & 0.087914 & 0.085358 & 0.092694 & 2048 & 2441 \\
8388608 & 0.1320806 & 0.123878 & 0.138708 & 4096 & 2048 \\
9998336 & 0.1456164 & 0.143222 & 0.147739 & 4096 & 2441 \\
16777216 & 0.2504316 & 0.235671 & 0.263854 & 4096 & 4096 \\
\end{tabular}
\caption{c-clang-o3-unix2-double2d/summary}\label{tab:res:c-clang-o3-unix2-double2d/summary}
\end{table}
\clearpage
\begin{table}
\centering
\begin{tabular}{llllll}
Pixels&Avg&Min&Max&Width&Height \\
\hline
10 & 0.0016884 & 0.001558 & 0.002037 & 5 & 2 \\
16 & 0.0017764 & 0.001607 & 0.002 & 4 & 4 \\
20 & 0.0016106 & 0.001486 & 0.00183 & 5 & 4 \\
32 & 0.0015504 & 0.001465 & 0.001622 & 8 & 4 \\
48 & 0.0016602 & 0.001563 & 0.001745 & 8 & 6 \\
64 & 0.0015994 & 0.001471 & 0.001777 & 8 & 8 \\
96 & 0.0016384 & 0.001504 & 0.001871 & 16 & 6 \\
128 & 0.001604 & 0.001497 & 0.001882 & 16 & 8 \\
192 & 0.0017834 & 0.001545 & 0.002014 & 16 & 12 \\
256 & 0.0016904 & 0.001511 & 0.001823 & 16 & 16 \\
480 & 0.0017136 & 0.001483 & 0.002092 & 32 & 15 \\
512 & 0.0016996 & 0.001621 & 0.001772 & 32 & 16 \\
992 & 0.0021332 & 0.001922 & 0.002458 & 32 & 31 \\
1024 & 0.0018952 & 0.00173 & 0.002183 & 32 & 32 \\
1984 & 0.001881 & 0.001633 & 0.002295 & 64 & 31 \\
2048 & 0.0020044 & 0.001722 & 0.002357 & 64 & 32 \\
4096 & 0.0020434 & 0.001925 & 0.002214 & 64 & 64 \\
4992 & 0.0020502 & 0.001826 & 0.002386 & 128 & 39 \\
8192 & 0.002019 & 0.001854 & 0.002192 & 128 & 64 \\
9984 & 0.002282 & 0.0021 & 0.002584 & 128 & 78 \\
16384 & 0.0022362 & 0.001974 & 0.002531 & 128 & 128 \\
19968 & 0.0023032 & 0.002066 & 0.002725 & 256 & 78 \\
32768 & 0.0022908 & 0.002146 & 0.002527 & 256 & 128 \\
49920 & 0.0025394 & 0.002454 & 0.00268 & 256 & 195 \\
65536 & 0.0026106 & 0.002465 & 0.002854 & 256 & 256 \\
99840 & 0.0028602 & 0.002692 & 0.003039 & 512 & 195 \\
131072 & 0.004063 & 0.003414 & 0.004472 & 512 & 256 \\
199680 & 0.0042526 & 0.003907 & 0.00493 & 512 & 390 \\
262144 & 0.0054336 & 0.004829 & 0.006129 & 512 & 512 \\
499712 & 0.010347 & 0.008566 & 0.012491 & 1024 & 488 \\
524288 & 0.0159684 & 0.010157 & 0.026783 & 1024 & 512 \\
999424 & 0.03524 & 0.03382 & 0.036304 & 1024 & 976 \\
1048576 & 0.0341834 & 0.033203 & 0.034891 & 1024 & 1024 \\
1998848 & 0.050911 & 0.048256 & 0.05331 & 2048 & 976 \\
2097152 & 0.0509186 & 0.048018 & 0.053225 & 2048 & 1024 \\
4194304 & 0.084699 & 0.082635 & 0.085924 & 2048 & 2048 \\
4999168 & 0.094597 & 0.090306 & 0.099331 & 2048 & 2441 \\
8388608 & 0.1519576 & 0.147526 & 0.155128 & 4096 & 2048 \\
9998336 & 0.1731964 & 0.167239 & 0.181819 & 4096 & 2441 \\
16777216 & 0.2775114 & 0.268871 & 0.288245 & 4096 & 4096 \\
\end{tabular}
\caption{c-gcc-o3-unix-double1d/summary}\label{tab:res:c-gcc-o3-unix-double1d/summary}
\end{table}
\clearpage
\begin{table}
\centering
\begin{tabular}{llllll}
Pixels&Avg&Min&Max&Width&Height \\
\hline
10 & 0.0015942 & 0.001495 & 0.001832 & 5 & 2 \\
16 & 0.0015684 & 0.001445 & 0.001764 & 4 & 4 \\
20 & 0.001737 & 0.001538 & 0.001903 & 5 & 4 \\
32 & 0.0016498 & 0.001496 & 0.001738 & 8 & 4 \\
48 & 0.0017384 & 0.001559 & 0.00198 & 8 & 6 \\
64 & 0.0015548 & 0.001451 & 0.001727 & 8 & 8 \\
96 & 0.0017344 & 0.001545 & 0.002038 & 16 & 6 \\
128 & 0.0017298 & 0.001614 & 0.001795 & 16 & 8 \\
192 & 0.001714 & 0.001647 & 0.001803 & 16 & 12 \\
256 & 0.001725 & 0.00162 & 0.00182 & 16 & 16 \\
480 & 0.0014872 & 0.001453 & 0.001528 & 32 & 15 \\
512 & 0.0014756 & 0.001431 & 0.001531 & 32 & 16 \\
992 & 0.0015144 & 0.001428 & 0.001541 & 32 & 31 \\
1024 & 0.0015218 & 0.001463 & 0.001655 & 32 & 32 \\
1984 & 0.0016216 & 0.001587 & 0.001667 & 64 & 31 \\
2048 & 0.0016046 & 0.001513 & 0.001696 & 64 & 32 \\
4096 & 0.0017422 & 0.00158 & 0.00182 & 64 & 64 \\
4992 & 0.0017202 & 0.001561 & 0.001809 & 128 & 39 \\
8192 & 0.0016064 & 0.001522 & 0.001676 & 128 & 64 \\
9984 & 0.0019246 & 0.001783 & 0.002209 & 128 & 78 \\
16384 & 0.0018426 & 0.001716 & 0.002076 & 128 & 128 \\
19968 & 0.001999 & 0.00185 & 0.002178 & 256 & 78 \\
32768 & 0.0019724 & 0.00186 & 0.002075 & 256 & 128 \\
49920 & 0.002186 & 0.001923 & 0.002377 & 256 & 195 \\
65536 & 0.00227 & 0.002155 & 0.002427 & 256 & 256 \\
99840 & 0.0026422 & 0.002539 & 0.002904 & 512 & 195 \\
131072 & 0.0028816 & 0.002597 & 0.003342 & 512 & 256 \\
199680 & 0.0032896 & 0.003106 & 0.003495 & 512 & 390 \\
262144 & 0.0042668 & 0.003888 & 0.004559 & 512 & 512 \\
499712 & 0.006514 & 0.005469 & 0.007185 & 1024 & 488 \\
524288 & 0.0171982 & 0.006816 & 0.02348 & 1024 & 512 \\
999424 & 0.0265278 & 0.025932 & 0.027361 & 1024 & 976 \\
1048576 & 0.0289088 & 0.02836 & 0.029143 & 1024 & 1024 \\
1998848 & 0.0387646 & 0.038396 & 0.038903 & 2048 & 976 \\
2097152 & 0.0400706 & 0.039558 & 0.04065 & 2048 & 1024 \\
4194304 & 0.0618162 & 0.060693 & 0.063196 & 2048 & 2048 \\
4999168 & 0.0715354 & 0.068445 & 0.076321 & 2048 & 2441 \\
8388608 & 0.116873 & 0.112857 & 0.125024 & 4096 & 2048 \\
9998336 & 0.1279968 & 0.12542 & 0.131373 & 4096 & 2441 \\
16777216 & 0.2004758 & 0.196424 & 0.208217 & 4096 & 4096 \\
\end{tabular}
\caption{c-gcc-o3-unix-int1d/summary}\label{tab:res:c-gcc-o3-unix-int1d/summary}
\end{table}
\clearpage
\begin{table}
\centering
\begin{tabular}{llllll}
Pixels&Avg&Min&Max&Width&Height \\
\hline
10 & 0.0047228 & 0.0043 & 0.005096 & 5 & 2 \\
16 & 0.0047542 & 0.004385 & 0.00529 & 4 & 4 \\
20 & 0.004802 & 0.004429 & 0.005216 & 5 & 4 \\
32 & 0.0051506 & 0.004829 & 0.005523 & 8 & 4 \\
48 & 0.0052322 & 0.004889 & 0.005583 & 8 & 6 \\
64 & 0.0053116 & 0.004893 & 0.006112 & 8 & 8 \\
96 & 0.0058564 & 0.005648 & 0.006216 & 16 & 6 \\
128 & 0.0063246 & 0.006185 & 0.006457 & 16 & 8 \\
192 & 0.0068208 & 0.006395 & 0.0074 & 16 & 12 \\
256 & 0.0072164 & 0.006806 & 0.007603 & 16 & 16 \\
480 & 0.0086172 & 0.007713 & 0.009607 & 32 & 15 \\
512 & 0.008345 & 0.007965 & 0.008634 & 32 & 16 \\
992 & 0.0118534 & 0.01142 & 0.012573 & 32 & 31 \\
1024 & 0.0119068 & 0.011527 & 0.012191 & 32 & 32 \\
1984 & 0.0177034 & 0.017344 & 0.018023 & 64 & 31 \\
2048 & 0.0180072 & 0.01768 & 0.018613 & 64 & 32 \\
4096 & 0.03073 & 0.030101 & 0.031897 & 64 & 64 \\
4992 & 0.0359412 & 0.035481 & 0.036261 & 128 & 39 \\
8192 & 0.0549972 & 0.05476 & 0.055377 & 128 & 64 \\
9984 & 0.0659984 & 0.065299 & 0.067556 & 128 & 78 \\
16384 & 0.1039074 & 0.103778 & 0.10401 & 128 & 128 \\
19968 & 0.1257842 & 0.125365 & 0.126259 & 256 & 78 \\
32768 & 0.2050054 & 0.202374 & 0.208976 & 256 & 128 \\
49920 & 0.3062182 & 0.305786 & 0.306879 & 256 & 195 \\
65536 & 0.399762 & 0.397993 & 0.401186 & 256 & 256 \\
99840 & 0.6054232 & 0.604796 & 0.606512 & 512 & 195 \\
131072 & 0.792189 & 0.791621 & 0.792473 & 512 & 256 \\
199680 & 1.2247622 & 1.222086 & 1.229689 & 512 & 390 \\
262144 & 1.6058994 & 1.596077 & 1.61638 & 512 & 512 \\
499712 & 3.050922 & 3.027957 & 3.088465 & 1024 & 488 \\
524288 & 3.18597 & 3.174444 & 3.219998 & 1024 & 512 \\
999424 & 6.0386108 & 6.032733 & 6.046526 & 1024 & 976 \\
1048576 & 6.3397604 & 6.328834 & 6.366426 & 1024 & 1024 \\
1998848 & 12.152556 & 12.081531 & 12.265659 & 2048 & 976 \\
2097152 & 12.6724314 & 12.631703 & 12.752535 & 2048 & 1024 \\
4194304 & 26.0175824 & 25.4087 & 27.014179 & 2048 & 2048 \\
4999168 & 30.1824428 & 30.140131 & 30.241431 & 2048 & 2441 \\
8388608 & 50.9606652 & 50.626181 & 51.547098 & 4096 & 2048 \\
9998336 & 60.3789196 & 60.270674 & 60.57919 & 4096 & 2441 \\
16777216 & 101.2790164 & 101.113425 & 101.63851 & 4096 & 4096 \\
\end{tabular}
\caption{hs/summary}\label{tab:res:hs/summary}
\end{table}
\clearpage






\end{document}